\documentclass{amsproc}

\usepackage{amsmath,mathtools} 
\usepackage{amsfonts}
\usepackage{xcolor}
\usepackage[hidelinks]{hyperref}
\usepackage{enumitem}

\usepackage{url}
\usepackage{tikz}
\usetikzlibrary{calc}

\newtheorem{theorem}{Theorem}[section]
\newtheorem{lemma}[theorem]{Lemma}

\theoremstyle{definition}
\newtheorem{definition}[theorem]{Definition}
\newtheorem{example}[theorem]{Example}
\newtheorem{proposition}[theorem]{Proposition}
\newtheorem{exercise}[theorem]{Exercise}

\theoremstyle{remark}

\numberwithin{equation}{section}

\DeclarePairedDelimiter{\SB}{[\![}{]\!]}
\DeclarePairedDelimiter{\RR}{\langle\!\langle}{\rangle\!\rangle}

\definecolor{green(ryb)}{rgb}{0.4, 0.69, 0.2}
\definecolor{frenchrose}{rgb}{0.96, 0.29, 0.54}
\definecolor{persianblue}{rgb}{0.11, 0.22, 0.73}
\definecolor{jade}{rgb}{0.0, 0.66, 0.42}
\definecolor{limegreen}{rgb}{0.2, 0.8, 0.2}

\begin{document}

\title{ $r$-matrices for integrable systems}


\author{}
\address{}
\curraddr{}
\email{}
\thanks{}

\author{Marta Dell'Atti}
\address{}
\curraddr{}
\email{m.dell-atti@uw.edu.pl}
\thanks{}


\date{}

\begin{abstract}
We consider some algebraic and geometric aspects of the theory of integrable systems in finite dimensions, associated with the existence of a classical $r$-matrix, first introduced by Sklyanin as the classical analogue of the quantum version. The importance of the notion of the $r$-matrix in this context relies on the fact that it connects the Hamiltonian structure of integrable equations with the factorisation problem which provides their explicit solution.

In this framework, the Lax matrix is interpreted as the coadjoint orbit of a Lie algebra $\mathfrak{g}$, and the existence of a non-dynamical $r$-matrix
allows the introduction of a second Lie algebra structure on $\mathfrak{g}$. Depending on the properties of the $r$-matrix associated with the specific
system, we distinguish between bialgebras and dialgebras. Bialgebras are associated with a skew-symmetric $r$-matrix, were introduced by Drinfeld, and connected to the interplay between the two Lie algebras structures on $\mathfrak{g}$ and its dual $\mathfrak{g}^*$ respectively. 
Dialgebras refer to a larger class of $r$-matrix and are related to the factorisation properties of the system, were introduced by Semenov-Tian-Shansky and consist in two Lie algebras $\mathfrak{g}$ and $\mathfrak{g}_R$ coexisting on the same vector space. 

\end{abstract}

\maketitle

\section*{Introduction} 
The theory of the classical $r$-matrix provides a beautiful description of integrable systems from a geometric and algebraic point of view, encompassing different features commonly shared by known integrable systems: 
\begin{enumerate}
    \item nonlinear integrable equations are compatibility conditions of an auxiliary system of linear equations;
    \item there exists a natural Poisson bracket with respect to which the systems are Hamiltonian;
    \item the integrals of motion are in involution with respect to the Poisson bracket and are spectral invariants of the auxiliary linear operator;
    \item the solution of the equations reduces to the Riemann-Hilbert problem in certain form. 
\end{enumerate}
The classical $r$-matrix was introduced by Sklyanin in \cite{Skl} as the classical version of the quantum $r$-matrix arising in the context of the research program initiated by Faddeev that lead to the quantum inverse scattering method and the theory of quantum groups~\cite{Fad}, whereas a geometric approach to study integrable systems finds its origin in the seminal works by Adler~\cite{Adl}, Kostant~\cite{Kos} and Symes~\cite{Sym}. Indeed, especially when the focus is on the splitting of the Lie algebra $\mathfrak{g}$, the method is referred to as Adler-Kostant-Symes scheme \cite{AdlerBook,BBT,KosLec}. 
The connection between the $r$-matrix and the geometric ideas is due to Semenov-Tian-Shansky~\cite{STS1,STS2}, who more recently introduced the notion of the Lie dialgebra~\cite{STSLec1,STSLec2} extending the concept of the Lie bialgebra coined by Drinfeld~\cite{Drinf1,Drinf2}. Even if the latter has been introduced earlier, it is much more instructive to start by studying the dialgebras and then the bialgebras. 

The aim of these lecture notes is to provide an overview of the theory of the $r$-matrix with an algebraic perspective to describe the integrable systems. We will deal with two objects $R\in\text{End}(\mathfrak{g})$ and $r\in\mathfrak{g} \otimes \mathfrak{g}$ that we will call $r$-matrix (and that are related under certain conditions). 
We will follow the treatment of Semenov-Tian-Shansky presented in~\cite{STSLec1} for dialgebras and that of Kosmann-Schwarzbach in \cite{KosLec} for bialgebras. The main other references are for the theory of integrable systems the books by Babelon, Bernard and Talon~\cite{BBT} and by Adler, van Moerbeke and Vanhaecke~\cite{AdlerBook}, and for the symplectic geometric aspects the reviews by Arnol'd, Givental' and Novikov~\cite{ArnGivNov} and by Butterfield~\cite{Butterfield}. Finally, we mention that the formalism of the $r$-matrix is currently implemented as a tool to determine features of integrable systems (\cite{BlaMar,CauDelSin,AZJ,FL}), or as a bridge connecting several different areas  (\cite{BHH,ZQ,H,Bertola}). 

The notes are structured as follows. 
 In Section~\ref{sec:geom_fund} we present the geometric approach leading to the identification in of the natural object to describe the kinematics: the coadjoint orbits in Poisson manifolds with respect to some Hamiltonian action of a Lie group $G$. We will define symplectic and Poisson manifolds, introduce symplectic and Hamiltonian actions of Lie groups, as well as the canonical projection and the momentum map. The interplay of the latter as a dual pair will be fundamental to identify the symplectic leaves of the Poisson manifold with the coadjoint orbits of the group. 
Section~\ref{sec:r_mat_dialgebra} is devoted to the introduction of the machinery for the $r$-matrix for dialgebras $(\mathfrak{g},\mathfrak{g}_R)$: here the $r$-matrix 
$R$ is an endomorphism of the Lie algebra $\mathfrak{g}$ and solution to the modified classical Yang-Baxter equation inducing a second Lie algebra structure $\mathfrak{g}_R$ acting on the same vector space. 
The dialgebra structure underpins a splitting of the Lie algebra $\mathfrak{g}$ into subalgebras. If the splitting is non-isotropic with respect to the pairing, one can always consider the dialgebra by building the double of the Lie algebra $\mathfrak{d}=\mathfrak{g}\oplus \mathfrak{g}$, and defining a feasible isotropic splitting there. An effect induced by the algebra splitting at the level of the Lie algebras is the factorisation at the level of the corresponding Lie group. 
Section~\ref{sec:r_mat_bialgebra} is dedicated to the study of the formalism of the $r$-matrix for bialgebras $(\mathfrak{g},\mathfrak{g}^*)$: here the $r$-matrix $r$ is an element of the tensor product of two copies of the Lie algebra $\mathfrak{g}$. On $\mathfrak{g}$ the $1$-cocycle map $\gamma$ is introduced, whose transpose defines the Lie bracket on $\mathfrak{g}^*$. The $r$-matrix enters at the level of maps as $\gamma=\delta r$, and the condition for $\gamma$ to endow $\mathfrak{g}^*$ with the Lie structure is that $r$ satisfies the modified classical Yang-Baxter equation. $r$ is here composed of an antisymmetric part and a symmetric part, and we will see under which constraints two formalism of $R$ and $r$ can be mapped one into the other. 
Lastly, in Section~\ref{sec:toda} we describe the paradigmatic example of the Toda chain, via both a dialgebra structure and a bialgebra structure. The dialgebra version is based on the classical relation between Toda and the Lie algebra of linear matrices, while the bialgebra structure is given in terms of the loop algebra for shift operators.


\section{Geometric fundamentals}\label{sec:geom_fund}   
The first question to address when dealing with the study of a mechanical system is its phase space, describing the kinematics. The phase space of a specific dynamical system is represented by a symplectic manifold, but as we will see the appropriate framework to deal with the Hamiltonian formalism is instead the Poisson manifold, i.e.\ a sort of a generalised version of the symplectic manifolds with some crucial differences. The Poisson manifold admits a stratification, whose layers --the symplectic leaves-- are the usual symplectic manifolds representing the phase spaces. The motivation for using this setting for the Hamiltonian formalism is well explained by Butterfield in~\cite{Butterfield} and it is three-fold: 
\begin{itemize}
    \item[$(i)$] A system of dimension $2n$ is naturally described by an even-dimensional space, but it may be useful to study the system at the varying of an odd number of parameters (say $s$) which are constant over time, but can distinguish different cases. The true dimension of the whole system is then $2n+s$ and if $s$ is odd, the natural structure is an odd-dimensional system. 
    \item[$(ii)$] Even in absence of any controllable parameters as in $(i)$, the description of several dynamical systems naturally leads to consider an odd-dimensional space. The case of the rigid body pivoted at a point is a paradigmatic example: its dynamics is expressed in terms of the three components of the angular momentum. 
    \item[$(iii)$] The configuration space of many dynamical systems is naturally taken to be a Lie group $G$, and the usual Hamiltonian mechanics is developed on the cotangent bundle $T^*G$. It is also natural to quotient this space by the lift to the cotangent bundle of the action of $G$ on itself, obtaining the reduced phase space $T^*G/G$, which is indeed a Poisson manifold. Moreover $T^*G/G \simeq \mathfrak{g}^*$ is isomorphic to the dual of the Lie algebra associated with $G$, and hence its symplectic leaves are the coadjoint orbits of $\mathfrak{g}^*$.
\end{itemize}
The first two reasons are related to the fact that the Poisson manifold can be degenerate, feature that will be crucial especially for the dialgebras. The last reason refers to the procedure known as Lie-Poisson reduction, leading to identify the natural setting for the Hamiltonian formalism: the coadjoint orbits on $\mathfrak{g}^*$.

\subsection{Poisson manifolds}

We introduce now two fundamental objects for this description: the Poisson manifold and the symplectic manifold. 

\begin{definition}
    A Poisson manifold $(\mathcal{M},\pi)$ is a manifold endowed with a Poisson bracket $\{\,\cdot\,,\,\cdot\,\}$, i.e.\ the Lie bracket on the space of smooth functions on $\mathcal{M}$ satisfying the Leibniz rule (on both entries)
\begin{equation}
    \{ \varphi, \psi\,\vartheta \} = \{ \varphi, \psi \} \,\vartheta +  \{ \varphi, \vartheta \} \,\psi\,, \qquad \varphi, \psi, \vartheta \in C^{\infty}(\mathcal{M})\,.
\end{equation}
\end{definition}
In local coordinates $\{x_i\}_{i=1}^n$ the bracket is expressed as 
\begin{equation}
    \{ \varphi, \psi\}(x) = \pi^{ij}(x) \, \frac{\partial \varphi}{\partial x_i}\, \frac{\partial \psi}{\partial x_j} \,, \qquad \varphi, \psi \in C^{\infty}(\mathcal{M})\,, 
\end{equation}
where the skew-symmetric tensor $\pi^{ij}$ is the so-called Poisson tensor. The Jacobi identity for the Poisson bracket is given by the system of equations
\begin{equation}
    \pi^{\ell i}\, \frac{\partial \pi^{jk}}{\partial x_{\ell}}+ \pi^{\ell j}\, \frac{\partial \pi^{ki}}{\partial x_{\ell}}+\pi^{\ell k}\, \frac{\partial \pi^{ij}}{\partial x_{\ell}} = 0\,, \qquad \text{for all } i,j,k\in \{ 1, \dots, n\}\,.
\end{equation}

Each function $\varphi$ on the Poisson manifold $\mathcal{M}$ induces a Hamiltonian vector field $\xi_{\varphi} \in \text{Vect}(\mathcal{M})$ defined by 
\begin{equation}
    \xi_{\varphi} = \{\varphi, \, \cdot\, \} \,, \qquad \xi_{\varphi} = \pi^{ij}(x) \, \frac{\partial \varphi}{\partial x_i}\, \frac{\partial }{\partial x_j}\,,
\end{equation}
with $\varphi$ the Hamiltonian of $\xi_{\varphi}$. 

\begin{definition}
A symplectic manifold $(\mathcal{S},\omega)$ is a manifold endowed with a closed non-degenerate $2$-form, and it is given with a natural Poisson structure. 
\end{definition}
In local coordinates the $2$-form is
\begin{equation}
    \omega = \omega_{ij}(x)\, dx^i \wedge dx^j\, \qquad i \neq j\,,
\end{equation}
with $\omega_{ij}$ non-degenerate. The closure of the $2$-form $d\omega=0$ is expressed as 
\begin{equation}
    \frac{\partial \omega_{jk}}{\partial x^i} +  \frac{\partial \omega_{ki}}{\partial x^j} + \frac{\partial \omega_{ij}}{\partial x^k} = 0\,, \qquad i \neq j \neq k\,. 
\end{equation}
Being the $2$-form non-degenerate, its inverse is well defined, and the natural Poisson bracket induced on the symplectic manifold $\mathcal{S}$ is 
\begin{equation}
    \{ \varphi, \psi \} = \omega_{ij}(x)^{-1} \, \frac{\partial \varphi}{\partial x_i} \, \frac{\partial \psi}{\partial x_j}\,,
\end{equation}
which satisfies the Jacobi identity. Vice versa, if the Poisson tensor $\pi^{ij}$ of a Poisson manifold $\mathcal{M}$ is non-degenerate induces a closed differential $2$-form 
\begin{equation}
    \omega = \pi^{ij}(x)^{-1}\, dx^i \wedge dx^j\,,
\end{equation}
hence equipping $\mathcal{M}$ with a symplectic structure\footnote{There is a crucial and profound difference between Poisson and symplectic manifolds in terms of their natural morphisms. Indeed, because of their definitions, the symplectic form transforms by the pullback (since it is a $2$-form), while the Poisson tensor by the pushforward (since it is a bi-vector). }. 

A result that dates back to Lie is that an arbitrary Poisson manifold (i.e.\ with no assumptions on the degeneracy of $\pi^{ij}$) admits a foliation: a stratification whose layers are the so-called symplectic leaves, each of which is a symplectic manifold. Geometrically, any function $H \in C^{\infty}(\mathcal{M})$ generates a Hamiltonian vector field $\xi_H$ acting on functions via the Poisson bracket
\begin{equation}
    \xi_H \cdot \varphi = \{H,\varphi \}\,, \qquad \text{for all }\varphi \in C^{\infty}(\mathcal{M})\,. 
\end{equation}
For any point of the manifold $x \in \mathcal{M}$ the tangent vectors $\xi_{H}(x)$ span a linear subspace in the tangent space $T_x\mathcal{M}$: this is the space tangent to the symplectic leaf passing through $x$. Since the Hamiltonian vector fields are tangent to the symplectic leaves, the Hamiltonian flows (i.e.\ $d\varphi/dt=\xi_H \cdot \varphi $) preserve each leaf separately.  

The property of a symplectic foliation characterising the Poisson manifold is related to the existence of nontrivial Casimir functions of the Poisson bracket. 
\begin{definition}
A Casimir function $\varphi$ is a function such that 
\begin{equation}
    \{ \varphi , \psi \} = 0 \,, \qquad \text{for all } \psi \in C^{\infty}(\mathcal{M})\,,
\end{equation}
i.e.\ it lives in the centre of the Poisson bracket.
\end{definition}
Therefore, the function $\varphi$ is a Casimir function if and only if it is constant on each symplectic leaf in~$\mathcal{M}$, giving rise to trivial equations of motion, an aspect that will be fundamental in Section~\ref{sec:r_mat_dialgebra}. On the other hand, level surfaces of Casimir functions induce a stratification on $\mathcal{M}$, that is in general less refined than that associated with the symplectic leaves, but provides a rather good description of the symplectic foliation. 

\subsection{Lie-Poisson brackets} In this section we will deal with the simplest example of a Poisson manifold that is not symplectic: $\mathfrak{g}^*$, the dual of a Lie algebra $\mathfrak{g}$. The natural Poisson structure defined on $\mathfrak{g}^*$ takes the name of Lie-Poisson structure\footnote{\label{foot:PL}Also called Berezin-Kirillov-Kostant-Souriau Poisson structure or a subset of these names~\cite{KosLec}. To avoid any confusion in the terminology, we specify that the Lie-Poisson structure is the Poisson bracket on $\mathfrak{g}^*$ mimicking the Lie bracket on $\mathfrak{g}$, while the Poisson-Lie group is a Lie group $G$ endowed with a quadratic Poisson structure, that we will introduce in Definition~\ref{PoissonLie}.}  
\begin{equation}\label{eq:LiePoisson_bracket}
    \{ X, Y \} (L) = \langle L\,, [X,Y] \rangle \,, \qquad X, Y \in \mathfrak{g}\,, \qquad L \in \mathfrak{g}^*\,, 
\end{equation}
where we identify with $\langle \, \cdot \,\,,\, \cdot \, \rangle$ the natural pairing between elements of $\mathfrak{g}$ and $\mathfrak{g}^*$. The Poisson bracket~\eqref{eq:LiePoisson_bracket} is an example of linear bracket, since it is linear with respect to elements of $\mathfrak{g}^*$, and for linear functions on $\mathfrak{g}$ it coincides with the Lie bracket on $\mathfrak{g}$. From linear functions (elements of $\mathfrak{g}$) the bracket can be extended to polynomials and then to arbitrary smooth functions, thanks to the Leibniz identity. Therefore, for functions $\varphi$, $\psi \in C^{\infty}(\mathfrak{g}^*)$ we have 
\begin{equation}\label{eq:Lie_PB}
    \{ \varphi, \psi\}(L) = \langle L\,, [d\varphi(L), d\psi(L)] \rangle \,, \qquad L \in \mathfrak{g}^*\,,
\end{equation}
and since $d\varphi(L), d\psi(L) \in (\mathfrak{g}^*)^* \simeq \mathfrak{g}$ the Lie bracket on the right hand side is well defined. 

The Poisson tensor $\pi^{ij}$ associated with the Lie-Poisson bracket is
\begin{equation}
    \pi^{ij}(L) = c^{ij}_k \, L^k \,, 
\end{equation}
with $c^{ij}_k$ structure constants of $\mathfrak{g}$ and $L^k = \langle L, e^{k}\rangle$ with $\{ e^k\}_{k=1}^n$ a basis in $\mathfrak{g}$. In local coordinates \eqref{eq:Lie_PB} becomes 
\begin{equation}
    \{ \varphi, \psi\}(L) = c^{ij}_k  L^k \, \frac{\partial \varphi(L)}{\partial L^i} \, \frac{\partial \psi(L)}{\partial L^j}\,. 
\end{equation}

The adjoint representation of the action of a Lie group $G$ on the Lie algebra $\mathfrak{g}$ is defined as 
\begin{equation}
    \text{Ad}_g X = \left. \frac{d}{dt}\,  g \cdot \text{exp}(t\,X) \cdot g^{-1} \right|_{t=0}\,, \qquad g \in G, X \in \mathfrak{g}\,. 
\end{equation}
If $G$ is a matrix Lie group, both $X$ and $g$ are matrices and we have the usual conjugation
\begin{equation}
    \text{Ad}_g\, X = g\,X\,g^{-1}\,. 
\end{equation}
The dual of this action is the coadjoint representation of the action of $G$ on its dual Lie algebra $\mathfrak{g}^*$, such as 
\begin{equation}
    \langle \text{Ad}^*_g\, L\,, X \rangle = \langle L\,, \text{Ad}_{g^{-1}}\, X\rangle\,, \qquad g\in G, \,  X \in \mathfrak{g}, \, L \in \mathfrak{g}^*\,. 
\end{equation}
Finally we have the (expected) adjoint and coadjoint representations of the action of the Lie algebra $\mathfrak{g}$ on itself, i.e.\ 
\begin{equation}
    \text{ad}_X\, Y = \left. \frac{d}{dt} \, \text{Ad}_{\text{exp}(tX)} Y\right|_{t=0}\,, \qquad \text{ad}^*_X\, Y = \left. \frac{d}{dt} \, \text{Ad}^*_{\text{exp}(tX)} Y\right|_{t=0}\,,
\end{equation}
hence
\begin{equation}
    \text{ad}_X Y = [X,Y]\,, \qquad (\text{ad}^*_X L) (Y) = - L(\text{ad}_X Y)\,. 
\end{equation}

\begin{definition} The coadjoint orbit of $L \in \mathfrak{g}^*$ is the set
\begin{equation}
    \mathcal{O}_L = \{ \text{Ad}^*_{g} L \,|\,g\in G \}\,,
\end{equation}
and the space tangent at $L \in \mathfrak{g}^*$ to the coadjoint orbit $\mathcal{O}_L$ is the linear subspace of $T_{L}\mathfrak{g}^*\simeq \mathfrak{g}^*$, i.e.\
\begin{equation}
    T_L \mathcal{O}_L = \{ \text{ad}^*_X L\,|\,X \in \mathfrak{g}\}\,. 
\end{equation}
\end{definition}

\begin{proposition}\label{prop:Ham_eq_Lie_Poisson}
Let $\varphi \in C^{\infty}(\mathfrak{g}^*)$ be an arbitrary function, then the Hamiltonian equation of motion defined by $\varphi$ with respect to the Lie-Poisson bracket can be expressed as 
\begin{equation}\label{eq:dl_tangent}
    \frac{dL}{dt} = - \text{ad}^*_{d\varphi(L)}L\,, \qquad L \in \mathfrak{g}^*\,.
\end{equation}
\end{proposition}
This means that the velocity vector for a Hamiltonian system on $\mathfrak{g}^*$ evaluated at some point $L \in \mathfrak{g}^*$ is tangent to the coadjoint orbit that goes through $L$. To show~\eqref{eq:dl_tangent}, we recall that associated with the function $\varphi$ there is the Hamiltonian vector field $\xi_\varphi$, and we consider an arbitrary element $X \in \mathfrak{g}$ and the linear function $X(L)=\langle L,X\rangle$. 
The Lie derivative\footnote{We recall that the connection between the Lie derivative and the Poisson bracket is the following~\cite{Butterfield}: the Lie derivative of $X$ along a vector field $\xi_{\varphi}$ is 
\[\mathcal{L}_{\xi_\varphi} X = dX(\xi_\varphi)=\xi_\varphi(X) = \{ \varphi, X\}\,. \]  } of $X(L)$ along $\xi_\varphi$ is 
\begin{equation}
    \begin{split} 
    \frac{dX(L)}{dt} &= \xi_\varphi X (L) = \{\varphi, X\}(L) = \langle L, [d\varphi(L),X]  \rangle \\
    &=  \langle L,\text{ad}_{d\varphi(L)}X  \rangle = - \langle \text{ad}^*_{d\varphi(L)}L, X \rangle\,.
    \end{split}
\end{equation}
Since $dX(L)/dt = \langle dL/dt,X \rangle$ and given the arbitrariness of $X$, we have~\eqref{eq:dl_tangent}. 

In many cases, the coadjoint orbits are indeed the phase spaces for a certain Lie group. Nevertheless, sometimes it will be useful to reduce the amount of degrees of freedom of an orbit, by taking the quotient on some symmetry group. Or in other cases, it is useful to consider a bigger phase space that is then mapped onto the orbit via a mapping preserving the Poisson structure. 

Now we consider two examples: the adjoint and coadjoint action of the Lie group $G$ on $\mathfrak{g}$ and $\mathfrak{g}^*$ coincide in the first example, and mismatch in the second example. We will often consider the matrix representation of Lie groups and algebras, and it will be useful to introduce the following inner product, i.e.\ the non-degenerate $\text{Ad}$-invariant bilinear form $\langle \,\cdot\,|\,\cdot\,\rangle$ that allows the identification $\mathfrak{g}^* \simeq \mathfrak{g}$
\begin{equation}\label{eq:inner_prod}
        \langle X\,|\,Y\rangle = \text{tr}(XY) \,, \qquad X,Y \in \mathfrak{g}\,. 
    \end{equation}

\begin{example}
    Let $\mathfrak{g}=\mathfrak{gl}(n,\mathbb{C})$ be the full matrix algebra with inner product~\eqref{eq:inner_prod}. In this case, the adjoint and coadjoint representations of the action of the Lie group coincide and the coadjoint orbits are 
    \begin{equation}
         \text{Ad}^*_g\,L =\text{Ad}_g\,L  = g\,L\,g^{-1} \,,\qquad L \in \mathfrak{g},\quad  g \in G\,,
    \end{equation}
    given by conjugated matrices, whose classification is possible in terms of their Jordan form. The orbit of $L \in \mathfrak{g}^*$ under the action of $G$ is then
    \begin{equation}
        \mathcal{O}_L = \{ g\,L\,g^{-1} \,|\,g \in G \}\,.
    \end{equation}
    The Casimir functions are the spectral invariants (i.e.\ the coefficients of the characteristic polynomial). 

\begin{exercise}
		Give a complete classification of coadjoint orbits for $G =GL(2,\mathbb{C})$ and for $G = GL(2,\mathbb{R})$. 
\end{exercise}
    
    
\end{example}


\begin{example}
    Let $\mathfrak{b}_+ \subset \mathfrak{g}$ be the algebra of upper triangular matrices, with inner product~\eqref{eq:inner_prod}. The related group $B_+$ is the group of invertible upper triangular matrices and the existence of the inner product inducing $\mathfrak{g}\simeq \mathfrak{g}^*$ allows us to determine the dual $\mathfrak{b}^*_+ \simeq \mathfrak{b}_-$ as the algebra of lower triangular matrices.     
    The adjoint and coadjoint action of~$B_+$ on $\mathfrak{b}_+$ and $\mathfrak{b}_-$ are
    \begin{align}
        \text{Ad}_b \,M &= b\, M\, b^{-1} \,, \qquad M \in \mathfrak{b}_+,~b \in B_+\,, \\[1ex]
        \text{Ad}_b^* \,M &= P_-( b\, M\, b^{-1} ) \,, \qquad M \in \mathfrak{b}_-,~b \in B_+\,,
    \end{align}
    with the projection $P_- \colon \mathfrak{g} \to \mathfrak{b}_-$. The orbit of $M$ under the action of $B_+$ on $\mathfrak{b_+^*}$ is 
    \begin{equation}
        \mathcal{O}_M = \{ P_-( b\, M\, b^{-1} ) \,|\,b\in B_+ \}\,. 
    \end{equation}
\end{example}
\begin{exercise}
	Describe all orbits of the group of upper triangular $2\times2$-matrices. 
\end{exercise}

The following important theorem due to Kostant and Kirillov holds true: 
\begin{theorem}
The symplectic leaves of the Lie-Poisson bracket are the coadjoint orbits of the coadjoint representation. The Casimir functions of the Lie-Poisson bracket are $\textup{Ad}^*\!$-invariant functions on $\mathfrak{g}^*$. 
\end{theorem}

The proof of this theorem will be the content of the next two sections.

\subsection{Hamiltonian Reduction}

To construct the symplectic leaves for a Poisson manifold, we will need to introduce the momentum map associated with the Hamiltonian action, and the concept of dual pairs of maps.

Let us consider the action of a Lie group $G$ on the Poisson manifold $\mathcal{M}$, i.e.\ $G \times \mathcal{M} \to \mathcal{M}$. 

\begin{definition} 
This of $G$ on $\mathcal{M}$ is said to be \emph{admissible} if the subspace $C^{\infty}(\mathcal{M})^G \subset C^{\infty}(\mathcal{M})$ of functions invariant with respect to the action of the group~$G$ is a Lie subalgebra with respect to the Poisson bracket, i.e.\ the Poisson bracket of $G$-invariant functions is itself $G$-invariant. 
\end{definition}

\begin{definition} 
A map $\alpha \colon \mathcal{M} \to \mathcal{N}$ with $\mathcal{M}$, $\mathcal{N}$ Poisson manifolds is known as Poisson map if it preserves the Poisson brackets, i.e.\
\begin{equation}
    \alpha^*\{ \varphi, \psi\}_{\mathcal{N}} = \{ \alpha^* \varphi, \alpha^* \psi \}_{\mathcal{M}}\,, \qquad \varphi, \psi \in C^{\infty}(\mathcal{N})\,,
\end{equation}
where $\alpha^* \varphi = \varphi \circ \alpha$. 
\end{definition} 

The action $G$ on the Poisson manifold $\mathcal{M}$ induces a natural Poisson map, as stated in the following.

\begin{proposition}
    Let the action $G \times \mathcal{M} \to \mathcal{M}$ be admissible and $\mathcal{M}/G$ be a smooth manifold. Then the space of orbits $\mathcal{M}/G$ is endowed with a natural Poisson structure such that the canonical projection map $\pi \colon \mathcal{M} \to \mathcal{M}/G $ is a Poisson map. 
\end{proposition}

The manifold $\mathcal{M}/G$ is said to be obtained by the reduction of $\mathcal{M}$ over the action of $G$. We are interested in two specific types of admissible actions of the group $G$: the symplectic action and the Hamiltonian action. 

\begin{definition} 
The action of the group $G$ on a symplectic manifold $\mathcal{S}$ is called \emph{symplectic action} if it preserves the symplectic structure, and this action is also admissible. 
\end{definition} 
An admissible symplectic action of the group $G$ on a Poisson manifold $\mathcal{M}$ means that the Poisson bracket itself is $G$-invariant (which is not true in general for admissible group actions)
\begin{equation}
    g^* \{ \varphi, \psi \} =\{g^* \varphi, g^* \psi  \} = \{ \varphi, \psi\}\,, \qquad \varphi, \psi \in C^{\infty}(\mathcal{M})^G,~g \in G\,. 
\end{equation}

Let us consider the action $G \times \mathcal{S} \to \mathcal{S}$ on the symplectic manifold $\mathcal{S}$. We also consider the homomorphism $\mathfrak{g} \to \text{Vect}(\mathcal{S})$ given by 
\begin{equation}\label{eq:vect_ham}
    \hat{X}_H(x) = \left. \frac{d}{dt} H\!\left(\exp(-tX) x\right) \right|_{t=t_0}\,, \qquad x \in \mathcal{S},~ X \in \mathfrak{g}\,.
\end{equation}
\begin{definition} 
The action of $G$ on $\mathcal{S}$ is called \emph{Hamiltonian action} if there is $H \colon \mathfrak{g} \to C^{\infty}(\mathcal{S})$ such that for any $X \in \mathfrak{g}$ the action of the one-parametric subgroup $\exp(-tX)$ is given by the Hamiltonian $H_X$. 
\end{definition} 
In other words, we can say that $\hat{X}_H$ is exactly the Hamiltonian vector field of $H_X$, and
thanks to the linearity of the map $X \mapsto H_X$ we have 
\begin{equation}\label{eq:Hamiltonian_action}
    H_{[X,Y]} = \{H_X, H_Y\}\,, \qquad H_X \in C^{\infty}(\mathcal{S}),~X,Y \in \mathfrak{g}\,.
\end{equation}
The importance of the Hamiltonian action is to be found in the nature of the Hamiltonian vector field, which being tangent to the leaves will induce the natural symplectic structure on the leaves of the stratification in form of a Hamiltonian action as well. The corresponding induced Poisson structure is then non-degenerate on any leaf.    

Symplectic actions are more general than Hamiltonian actions. In particular, for a symplectic action, in presence of all vector fields globally Hamiltonian generated by the action of $G$, the relation~\eqref{eq:Hamiltonian_action} becomes 
\begin{equation}
    H_{[X,Y]} - \{H_X, H_Y\} = c(X,Y) = \text{const}\,,
\end{equation}
since the constant functions are in the centre of the Poisson bracket. One can see that $c(X,Y)$ is a skew bilinear form on $\mathfrak{g}$ satisfying
\begin{equation}
    c([X,Y],Z) + c([Y,Z],X) + c([Z,X],Y) = 0\,,
\end{equation}
i.e.\ we can say that it is a $2$-cocycle\footnote{Cocycles, coboundaries and cochains will be properly introduced in Section~\ref{sec:r_mat_bialgebra}.}. Also, with $\hat{\mathfrak{g}}$ the central extension of $\mathfrak{g}$ corresponding to $c$, the symplectic action of $G$ becomes Hamiltonian with respect to $\hat{\mathfrak{g}}$. 


An important element to construct the symplectic leaves is given by the momentum map of a Lie group action $G$ on the Poisson manifold $\mathcal{M}$. The momentum map represents the geometric generalisation of a conserved quantity (i.e.\ the linear momentum for the Euclidean group). 

\begin{definition}
The momentum map is a map $\mu \colon \mathcal{M} \to \mathfrak{g}^*$ with values on the dual of the Lie algebra of the group $G$. With $H$ the Hamiltonian invariant with respect to  the action of the Lie group $G$,  the momentum map is 
\begin{equation}
    \mu(x)(X) = H_X(x) \,, \qquad x \in \mathcal{M},\quad H \in C^{\infty}(M)^G \,.
\end{equation}
\end{definition}
The conservation of the momentum maps allows us to define a geometric version of the Noether's theorem. We now consider two examples of momentum maps. 

\begin{example}
    Consider the group of translations $G=
    \mathbb{R}^3$ of a free particle, hence acting on the manifold $\mathcal{M}=\mathbb{R}^3$, which represents the particle's configuration space. For $T^*(\mathbb{R}^3)\simeq \mathbb{R}^3 \times \mathbb{R}^3$. The Hamiltonian action is 
    \begin{equation}
        (q,p) \mapsto (q+x,p)\,, \qquad x \in \mathbb{R}^3\,,
    \end{equation}
    and the corresponding momentum map  is 
    \begin{equation}
        \mu \colon T^*(\mathbb{R}^3)\to \mathbb{R}^3 \,, \qquad \langle \mu(q,p), x \cdot (e_1,e_2,e_3) \rangle = p \cdot x\,, 
    \end{equation}
    and fixing the basis, we have the usual linear momentum, i.e.\ $\mu(q,p)=p$. 
\end{example}

\begin{example}
    Consider the action of $G=SO(3)$ on $\mathcal{M}=\mathbb{R}^3$, which induces a Hamiltonian action on $T^*(\mathbb{R}^3) \cong \mathbb{R}^3 \times \mathbb{R}^3$ via
\begin{equation}
(q, p) \mapsto (Aq, pA^{-1})\,, \quad A \in SO(3)\,.
\end{equation} 
Then the momentum map for this Hamiltonian action is
\begin{equation}
\mu : T^*(\mathbb{R}^3) \to \mathfrak{so}(3)^*, \quad \langle \mu(q, p), X \cdot (\Omega_1, \Omega_2, \Omega_3) \rangle = (q \wedge p) \cdot X
\end{equation}
where
\[
\Omega_1 = \begin{pmatrix}
0 & 0 & 0 \\
0 & 0 & -1 \\
0 & 1 & 0
\end{pmatrix}, \quad
\Omega_2 = \begin{pmatrix}
0 & 0 & 1 \\
0 & 0 & 0 \\
-1 & 0 & 0
\end{pmatrix}, \quad
\Omega_3 = \begin{pmatrix}
0 & -1 & 0 \\
1 & 0 & 0 \\
0 & 0 & 0
\end{pmatrix}.
\]
With $\Omega_1, \Omega_2, \Omega_3$ as an orthonormal basis of $\mathfrak{so}(3)$, and further identifying $\mathfrak{so}(3) \cong \mathfrak{so}(3)^* \cong \mathbb{R}^3$, we find the usual angular momentum, i.e.\ $\mu(q, p) = q \wedge p$. 
\end{example}

The momentum map associated with the Hamiltonian action of the Lie group $G$ is a Poisson map because because by construction it is an equivariant map, i.e.\ the following holds. 

\begin{proposition}
    Let $\mathfrak{g}^*$ be the dual of a Lie algebra endowed with a Lie-Poisson bracket. The following diagram is commutative: 
    \begin{equation*}
    \centering
    \begin{tikzpicture}
        \path (0,0) node (a) {$G \times \mathcal{M}$} -- ++(2,0) node (b) {$\mathcal{M}$} -- ++(0,-2) node (c) {$\mathfrak{g}^*$} -- ++(-2,0) node (d) {$G \times \mathfrak{g}^*$};  
        \draw[->] (a) -- (b);
        \draw[->] (d) -- (c) node[above,midway] (f) {\small$\text{Ad}^*$};
        \draw[->] (a) -- (d) node[left,midway] (f) {\small$\text{id} \times \mu$};
        \draw[->] (b) -- (c) node[right,midway] (f) {\small$\mu$};
    \end{tikzpicture}
    \end{equation*}
with vertical arrows Poisson maps. Vice versa, any Poisson map $\mu \colon \mathcal{M} \to \mathfrak{g}^*$ induces a Hamiltonian action of $G$ on $\mathcal{M}$ for which $\mu$ is the associated momentum map. The fact that the Hamiltonian vector field is by construction tangent to the symplectic leaves, the Hamiltonian action induces a Hamiltonian action on every symplectic leaf, in the sense of symplectic geometry. 
\end{proposition}


A fundamental concept in the theory of Poisson manifolds is that the momentum map $\mu \colon \mathcal{M} \to \mathfrak{g}^*$ and the canonical projection $\pi \colon \mathcal{M} \to \mathcal{M}/G$ form a \emph{dual pair} of Poisson maps. We now define the dual pair in general and then we will specify it for the case of interest. Let $\mathcal{N}_1, \mathcal{N}_2$ and $\mathcal{M}$ be Poisson manifolds with $\mathcal{N}_i \subset \mathcal{M}$ for $i=1,2$ and $\alpha_i \colon \mathcal{M} \to \mathcal{N}_i$ for $i=1,2$ two Poisson maps. Let $\alpha_i^*C^{\infty}(\mathcal{N}_i)$ be the subalgebra of smooth functions $C^{\infty}(\mathcal{M})$ that are the pullback of $\mathcal{N}_i$ for $i=1,2$. 

\begin{definition}
Two Poisson maps $\alpha_1$, $\alpha_2$ form a dual pair if the subalgebra $\alpha_1^*C^{\infty}(\mathcal{N}_1)$ is the {centraliser} in $C^{\infty}(\mathcal{M})$ of $\alpha_2^*C^{\infty}(\mathcal{N}_2)$ with respect to the Poisson bracket on $\mathcal{M}$ (and the same for the exchange $1 \leftrightarrow 2$). 
\end{definition}

Then the following theorem holds. 

\begin{theorem}
    Let 
    \begin{equation*}
        \begin{tikzpicture}
            \path (0,0) node (a) {$\mathcal{M}$} -- ++(-1,-1.2) node (b) {$\mathcal{N}_1$} -- ++(2,0) node (c) {$\mathcal{N}_2$};
            \draw[->] (a) -- (b) node[midway,left] {\small$\alpha_1$}; 
            \draw[->] (a) -- (c) node[midway,right] {\small$\alpha_2$}; 
        \end{tikzpicture}
    \end{equation*}
    be a dual pair. Then, the connected components of the sets $\alpha_1(\alpha_2^{-1}(v))$, $v \in \mathcal{N}_2$ and $\alpha_2(\alpha_1^{-1}(u))$, $u \in \mathcal{N}_1$ are symplectic leaves in $\mathcal{N}_1$ and $\mathcal{N}_2$ respectively. 
\end{theorem}

For Hamiltonian actions of the group $G$ on the Poisson manifold $\mathcal{M}$ then we have the following. 

\begin{proposition}
    Let $G \times \mathcal{M}$ be a Hamiltonian action of the group $G$, the map $\pi \colon \mathcal{M} \to \mathcal{M}/G$ be the canonical projection to the $G$-orbits space, and the map $\mu \colon \mathcal{M} \to \mathfrak{g}^*$ be the associated momentum map. Then 
    \begin{equation*}
        \begin{tikzpicture}
            \path (0,0) node (a) {$\mathcal{M}$} -- ++(-1,-1.2) node (b) {$\mathcal{M}/G$} -- ++(2,0) node (c) {$\mathfrak{g}^*$};
            \draw[->] (a) -- (b) node[midway,left] {\small $\pi$}; 
            \draw[->] (a) -- (c) node[midway,right] {\small $\mu$}; 
        \end{tikzpicture}
    \end{equation*}
    is a dual pair. As a consequence, the symplectic leaves in $\mathcal{M}/G$ are connected components of the sets $\pi(\mu^{-1}(L))$, for $L \in \mathfrak{g}^*$. 
\end{proposition}

It is important to specify that the canonical projection $\pi$ being a quotient map is not a homomorphism. Indeed, it does not preserve several properties of the manifold $\mathcal{M}$ to the quotient manifold $\mathcal{M}/G$, but the presence of connected components in $\mathcal{M}$ is preserved in $\mathcal{M}/G$ under its action. In particular, a quotient map $\pi \colon \mathcal{M} \to \mathcal{M}/G$ is such if it is surjective, continuous and the pre-image $\pi^{-1}(\mathcal{U})\in \mathcal{M}$ being an open set implies that $\mathcal{U} \in \mathcal{M}/G$ is open. 

In the following we consider two examples of quotient maps built starting from different equivalence relations, to give a visualisation of how different the quotiented space can be compared to the initial one. 

\begin{example}
    We consider the one-dimensional manifold $(\mathbb{R}, \mathcal{T}_{\mathbb{R}})$, where with $\mathcal{T}_{\mathbb{R}}$ we identify the topology of the manifold $\mathbb{R}$. Let $\pi \colon \mathbb{R} \to \mathbb{R}/\!\sim$ be the canonical projection with respect to the equivalence relation $\sim$ such that 
    \begin{equation}
        x \sim y, \quad  y = x + N\,, \quad N \in \mathbb{Z}\,,
    \end{equation}
    so that the final manifold is $\mathbb{R}/\mathbb{Z}$. One can show that the map $\alpha \colon \mathbb{R} \to S^1$ mapping the real axis on the circle $(S^1, \mathcal{T}_{S^1})$ is as well a quotient map and $\mathbb{R}/\mathbb{Z} \simeq S^1$ are homeomorphic. 
    
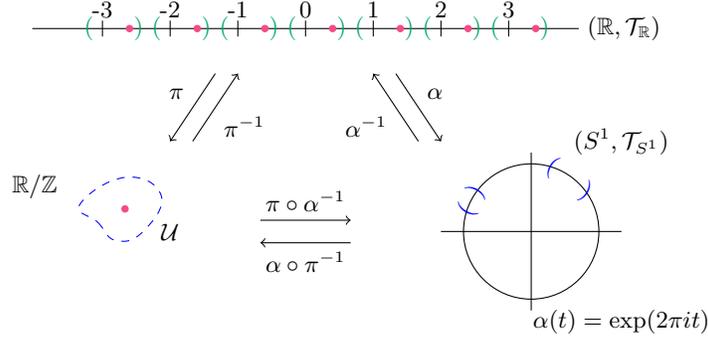
\begin{equation*}
        \begin{tikzpicture}[scale=.6]\small
        \draw ($(-5,.5)+(-30pt,0)$) -- ($(5,.5)+(30pt,0)$) node[right] {$(\mathbb{R},\mathcal{T}_{\mathbb{R}})$};
        \foreach \x in {-3,-2,-1,...,3}
    {        
      \coordinate (A\x) at ($(0,.5)+(\x*1.5cm,0)$) {};
      \coordinate (B\x) at ($(.6cm,.5)+(\x*1.5cm,0)$) {};
      \node[jade,thick] (C\x) at ($(-.3cm,.5)+(\x*1.5cm,0)$) {$($};
      \node[jade,thick] (D\x) at ($(.8cm,.5)+(\x*1.5cm,0)$) {$)$};
      \draw ($(A\x)+(0,5pt)$) -- ($(A\x)-(0,5pt)$);
      \node at ($(A\x)+(0,3ex)$) {\x};
      \filldraw[frenchrose] (B\x) circle (2pt);
    }
    \draw[->,] (-2,-.5) -- (-3,-2) node[midway,above left] {\small$\pi$};
    \draw[->,] (-2.5,-2) -- (-1.5,-.5) node[midway,below right] {\small$\pi^{-1}$};
    \draw[->,] (2,-.5) -- (3,-2) node[midway,above right] {\small$\alpha$};
    \draw[->,] (2.5,-2) -- (1.5,-.5) node[midway,below left] {\small$\alpha^{-1}$};
    \node (a) at  (-6,-3) {$\mathbb{R}/\mathbb{Z}$};
     \filldraw[frenchrose] (-4,-3.5) circle (2pt);
     \begin{scope}[scale=.4]
     \draw[dashed,blue] plot [smooth cycle, tension=.8] coordinates {($(-.5,1.5)+(-12,-10)$)($(.5,.5)+(-12,-10)$)($(1.5,-.5)+(-12,-10)$)($(3,0)+(-12,-10)$)($(4,2)+(-12,-10)$)($(2,3)+(-12,-10)$)};
     \end{scope}
     \node (c) at (-3,-4) {$\mathcal{U}$};
     \begin{scope}
     \draw (5,-4) circle (1.5cm);
     \node[blue,rotate=-30] (A) at ($(5,-4) +(0.2*180:1.5cm)$) {$)$};
     \node[blue,rotate=-30] (A) at ($ (5,-4) +(0.4*180:1.5cm)$) {$($};
     \node[blue,rotate=60] (A) at ($ (5,-4) +(0.8*180:1.5cm)$) {$)$};
     \node[blue,rotate=60] (A) at ($ (5,-4) +(0.9*180:1.5cm)$) {$($};
     \draw (5,-2.25) -- (5,-5.75);
     \draw (3,-4) -- (7,-4);
     \node (c1) at (7,-2) {$(S^1,\mathcal{T}_{S^1})$};
     \node (c2) at (7,-6) {$\alpha(t)=\exp(2\pi i t)$};
     \draw[->] (-1,-3.75) -- (1,-3.75) node[midway,above] {\small $\pi \circ \alpha^{-1}$};
    \draw[->] (1,-4.25) -- (-1,-4.25) node[midway,below] {\small$\alpha \circ \pi^{-1}$};
     \end{scope} 
        \end{tikzpicture}    
        \end{equation*}
        Hence, connected components of $(\mathbb{R}, \mathcal{T}_{\mathbb{R}})$ are mapped into connected components of $(\mathbb{S^1}, \mathcal{T}_{S^1})$. 
\end{example}

\begin{example}
    We consider the quotient map $\alpha \colon \mathbb{R} \to \mathbb{R}/\!\sim$ with the equivalence relation 
    \begin{equation}
        x \sim y \,,\quad (x,y) \in \mathbb{Z}_2 \,,
    \end{equation}
    meaning that $x$ and $y$ are equivalent if both are integers, 
    while for all the other points of $\mathbb{R}$ only the reflexive property is true, i.e.\ $x \sim x$ for $x \notin \mathbb{Z}$. 
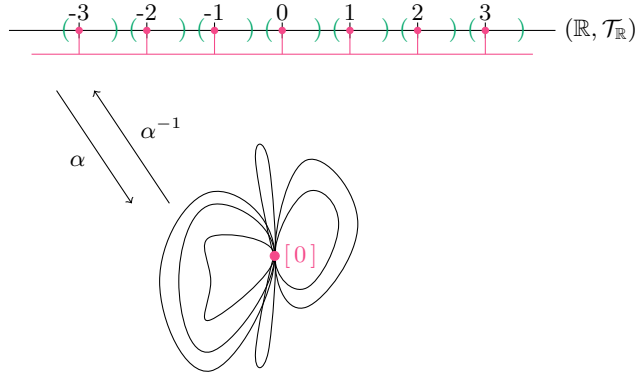
    \begin{equation*}
        \begin{tikzpicture}\small 
        \begin{scope}[scale=.6]
           \draw ($(-5,.5)+(-30pt,0)$) -- ($(5,.5)+(30pt,0)$) node[right] {$(\mathbb{R},\mathcal{T}_{\mathbb{R}})$};
        \foreach \x in {-3,-2,-1,...,3}
    {        
      \coordinate (A\x) at ($(0,.5)+(\x*1.5cm,0)$) {};
      \coordinate (B\x) at ($(.6cm,.5)+(\x*1.5cm,0)$) {};
      \node[jade,thick] (C\x) at ($(-.3cm,.5)+(\x*1.5cm,0)$) {$($};
      \node[jade,thick] (D\x) at ($(.8cm,.5)+(\x*1.5cm,0)$) {$)$};
      \draw ($(A\x)+(0,5pt)$) -- ($(A\x)-(0,5pt)$);
      \draw[frenchrose] ($(A\x)$) -- ($(A\x)-(0,15pt)$);
      \node at ($(A\x)+(0,3ex)$) {\x};
      \filldraw[frenchrose] (A\x) circle (2pt);
    }
    \draw[frenchrose] ($(-4.5,.5)+(-30pt,-15pt)$) -- ($(4.5,.5)+(30pt,-15pt)$);
    \end{scope}
    \draw[->,] (-3,-.5) -- (-2,-2) node[midway,below left] {\small$\alpha$};
    \draw[->,] (-1.5,-2) -- (-2.5,-.5) node[midway,above right] {\small$\alpha^{-1}$};
    \coordinate (c0) at ($(A1)+(-1,-3cm)$);
    \begin{scope}[scale=.85]
    \draw plot [smooth cycle, tension=.8] coordinates {($(c0)$)($(c0)+(-1,-1)$)($(c0)+(-1,-0.4)$)($(c0)+(-1,.3)$)};
    \draw plot [smooth cycle, tension=.8] coordinates {($(c0)$)($(c0)+(-1,-1.5)$)($(c0)+(-1.5,-0.4)$)($(c0)+(-1,.8)$)};
    \draw plot [smooth cycle, tension=.8] coordinates {($(c0)$)($(c0)+(-1,-1.8)$)($(c0)+(-1.8,-0.4)$)($(c0)+(-1,1)$)};
    \draw plot [smooth cycle, tension=.8] coordinates {($(c0)$)($(c0)+(.5,1)$)($(c0)+(1,.5)$)($(c0)+(.5,-.5)$)};
    \draw plot [smooth cycle, tension=.8] coordinates {($(c0)$)($(c0)+(.5,1.5)$)($(c0)+(1.3,.5)$)($(c0)+(.5,-.8)$)};
    \draw plot [smooth cycle, tension=.8] coordinates {($(c0)$)($(c0)+(-.3,1.5)$)($(c0)+(-.1,1.5)$)};
    \draw plot [smooth cycle, tension=.8] coordinates {($(c0)$)($(c0)+(-.3,-1.5)$)($(c0)+(-.1,-1.5)$)};
    \filldraw[frenchrose] (c0) circle (2pt) node[right] {$[\,0\,]$};
    \end{scope} 
        \end{tikzpicture}
    \end{equation*}
    Again, the connected components are preserved after the action of the quotient map. 
\end{example}

The action of the group $G$ on the manifold $\mathcal{M}$ induces the canonical map $\pi \colon \mathcal{M} \to \mathcal{M}/G$, which defines a reduction. 
A Hamiltonian $H$ invariant with respect to the action of the group $G$ on the Poisson manifold $\mathcal{M}$ gives rise to a reduced Hamiltonian $\bar{H}$ on $\mathcal{M}/G$. By definition, $\bar{H}$ is a function on $\mathcal{M}/G$ such that $H = \bar{H} \circ \pi$ (from $\bar{H}=\pi^*H$). Therefore, the canonical projection $\pi$ maps integral curves of $H$ onto the integral curves of $\bar{H}$. Let $F_t$ be the dynamical flow on $\mathcal{M}$ defined by $H$, $\bar{F}_t$ by $\bar{H}$. Then the diagram 
\begin{equation*}
    \begin{tikzpicture}
        \path (0,0) node (a) {$\mathcal{M}$} -- ++(2,0) node (b) {$\mathcal{M}$} -- ++(0,-2) node (c) {$\mathcal{M}/G$} -- ++(-2,0) node (d) {$\mathcal{M}/G$};  
        \draw[->] (a) -- (b) node[above,midway] (f) {\small$F_t$};
        \draw[->] (d) -- (c) node[above,midway] (f) {\small$\bar{F}_t$};
        \draw[->] (a) -- (d) node[left,midway] (f) {\small$\pi$};
        \draw[->] (b) -- (c) node[right,midway] (f) {\small$\pi$};
    \end{tikzpicture}
\end{equation*}
is commutative and $F_t$ is said to \emph{factorise} over $\mathcal{M}/G$. 

We will apply this construction to the cotangent bundle of a Lie group. 

\subsection{Cotangent bundle of a Lie group}
Let $\mathcal{M}$ be a smooth manifold and $T^*\mathcal{M}$ its cotangent bundle. We now recall the notion of the symplectic structure defined on $T^*\mathcal{M}$. With $q \in \mathcal{M}$ and $p \in T_q^*\mathcal{M}$ we have that the tangent space to $T^*\mathcal{M}$ can be seen as 
\begin{equation}
    T_{(q,p)}(T^*\mathcal{M}) \simeq T_q \mathcal{M} \oplus T_q^*\mathcal{M} \,.
\end{equation}
Let $\alpha \colon T_{(q,p)}(T^*\mathcal{M}) \to T_q\mathcal{M}$ be the projection in this splitting. The canonical $1$-form $\theta$ and the canonical symplectic form $\omega$ (both on $T^*\mathcal{M}$) are 
\begin{equation}
    \theta(v) = \langle p\,, \alpha(v) \rangle \,, \qquad \omega = d\theta\,, \qquad v \in T_{(q,p)}(T^*\mathcal{M})\,, 
\end{equation}
and in local coordinates, these take the standard form 
\begin{equation}
    \theta = p_i \, dq^i\,, \qquad \omega = dp_i \wedge dq^i\,. 
\end{equation}

Let $G$ be a Lie group and $T^*G$ its cotangent bundle. The action of $G$ on itself by left and right translation is free, therefore the tangent and cotangent bundle are trivial, i.e.\ these actions can be expressed as Cartesian products. Let $\lambda_x$ be the left translation map, i.e.\ 
\begin{equation}
\begin{split} 
    \lambda_x &\colon G \to G\,,\qquad x \in G\,, \\[1ex]
    \lambda_x &\colon g \mapsto xg\,, 
\end{split} 
\end{equation}
$\lambda_x'$ its differential at the unit element $e$ and its dual $\lambda_x'^*$, given by
\begin{equation}
    \begin{split}
        \lambda_x'&\colon T_eG \simeq \mathfrak{g} \to T_x G\,, \\[1ex]
        \lambda_x'^*&\colon T_x^*G \to T_e^* G \simeq \mathfrak{g}^*\,. 
    \end{split}
\end{equation}
A different trivialisation of $T^*G$ exploits the differentials of the right translation~$\rho_x$. In left trivialisation, the action of $G$ on the cotangent bundle $T^*G\simeq G \times \mathfrak{g}^*$ is 
\begin{equation}\label{eq:lambda_rho}
    \begin{split}
        \lambda_x &\colon ( g \,,\, L) \mapsto (xg \,,\, L)\,, \\[1ex]
        \rho_x &\colon (g \,,\, L) \mapsto (gx^{-1}\,,\, \text{Ad}^*_x L )\,, 
    \end{split}
\end{equation}
since fixing the left action $\lambda_{x}$, the right action $\rho_{x}\colon g \mapsto gx$ is obtained by inversion 
\begin{equation}
    \rho_x = \lambda_x^{-1} = \lambda_{x^{-1}}\,, \quad x \in G\,.
\end{equation}

The left and right actions $\lambda$, $\rho$ are defined on $G \times G \to G$, which admit a cotangent lift to $G \times T^*G \to T^*G$ preserving the symplectic structure, and furthermore they are Hamiltonian. Therefore, they also admit momentum maps. In particular, the momentum map  $\mu_{\lambda} \colon T^*G \to \mathfrak{g}^*$ of the cotangent lift of $\lambda$ is the cotangent lift of the right translation, and analogously the momentum map  $\mu_{\rho} \colon T^*G \to \mathfrak{g}^*$ of the cotangent lift of $\rho$ is the cotangent lift of the left translation. In left trivialisation, they are then 
\begin{equation}\label{eq:moments}
        \begin{split}
            \mu_{\lambda} \colon (x\,, L) &\mapsto - \text{Ad}^*_x L \,, \\[1ex] 
            \mu_{\rho} \colon (x\,, L) &\mapsto L\,.
        \end{split}
    \end{equation}
We emphasise that the sign appearing in $\mu_{\lambda}$ depends on the choice of the definition of the Hamiltonian vector field~\eqref{eq:vect_ham} and it is also responsible for the sign in the Hamiltonian action~\eqref{eq:Hamiltonian_action}. This allows us to consider that the quotient space $T^*G/G$ under the canonical projection under the left action is isomorphic to $\mathfrak{g}^*$ endowed with the Lie-Poisson bracket. The corresponding quotient space under the canonical projection induced by the action for right translations is instead anti-isomorphic to $\mathfrak{g}^*$ (i.e.\ the Lie-Poisson bracket changes sign). 

The maps $\mu_{\lambda}$, $\mu_{\rho}$ form a dual pair in $T^*G$, and in particular $\mu_{\lambda}$ is constant on orbits of $\rho$, and $\mu_{\rho}$ is constant on orbits of $\lambda$. The form of the lifted Poisson bracket defined on functions on $T^*G$ in left trivialisation is:\footnote{The Poisson bracket can be constructed by considering the inverse of the symplectic form $\omega=d\theta$ on $T^*G$, where $\theta(g,L)=\langle L, \omega_{\text{MC}} \rangle$ and $\omega_{\text{MC}}\in \mathfrak{g}$ the Maurer-Cartan form. The latter is defined in terms of left-invariant $1$-forms on $G$ corresponding to $e^i \in \mathfrak{g}^*$
\begin{equation}
    \omega_{\text{MC}} = e_i \otimes \omega^i \,, 
\end{equation}
and for $G$ Lie group one has $\omega_{\text{MC}}= g^{-1}\, dg$. 
} 
\begin{equation}
    \{ \varphi, \psi\}(g,L) = \partial^i \varphi\, \hat{e}_i \psi - \partial^i \psi\, \hat{e}_i \varphi + \frac{1}{2}\, c^k_{ij} L_k\, \partial^i \varphi\, \partial^j \psi\,, 
\end{equation}
where $\partial^i=\partial/\partial L^i$, $\{e_i\}_{i=1}^n$ a basis for the Lie algebra $\mathfrak{g}$, $\{e^i\}_{i=1}^n$ for $\mathfrak{g}^*$ such that $L = L_i\, e^i$, and the left invariant field on $G$ generated by
\begin{equation}
    \hat{X} \, \varphi(x) = \left.\frac{d}{dt} \varphi(x\,\exp(tX)) \right|_{t=0}\,, \qquad X \in \mathfrak{g}\,.  
\end{equation}

\begin{proposition}
The symplectic leaves of $\mathfrak{g}^*$ coincide with the orbits of the coadjoint representation. 
\end{proposition}
Indeed, we have from~\eqref{eq:moments} 
\begin{equation}
    \mu_{\lambda}(\lambda^{-1}(L))=  \mu_{\lambda}(\mu_{\rho}^{-1}(L))= \mu_{\lambda}(\mu_{\rho}(L))=\{-\text{Ad}^*_g L\,, g \in G\} = \mathcal{O}_{-L}\,, 
\end{equation}
establishing that the most convenient framework for the kinematics and the description of the phase space is provided by the coadjoint orbits of $\mathfrak{g}^*$. 

After the identification of the coadjoint orbits as the natural setting to describe phase spaces for the kinematics, we will present the Hamiltonian formulation for the dynamics of integrable systems. We will introduce
two fundamental algebra structures (dialgebras and bialgebras), related to two different notions of the so-called $r$-matrices: as a map $R \colon \mathfrak{g} \to \mathfrak{g}$ in the framework of linear Poisson structures, and as the map $r \colon \mathfrak{g} \otimes \mathfrak{g} \to \mathfrak{g}$  for nonlinear Poisson structures.

\section{\boldmath \texorpdfstring{$r$-matrices for Lie dialgebras}{rMat}}\label{sec:r_mat_dialgebra}

The modern theory of integrable systems takes its origin on the fundamental observation due to Lax in 1967 that it is possible to write a nonlinear integrable equation in Lax form, i.e.\
\begin{equation}\label{eq:Lax_eq}
    \frac{dL}{dt} = [M,L]\,, 
\end{equation}
where $L$, $M$ are operators depending on the dynamical variables. Usually, $L$ contains information about the initial data, while $M = M(L)$, i.e.\ it is a function of the entries of $L$. The integral of motion for this evolutionary equation are the spectral invariants of the operator $L$. Faddeev and Zakharov observed in~\cite{FedZak} by considering the example of the KdV equation, that Lax equations are Hamiltonian and the spectral invariants of $L$ form a complete set of integrals in evolution. 

Even if the Hamiltonian aspects were known in \cite{FedZak}, the way to provide an appropriate description of the Hamiltonian structure in relation to the commutation appearing in~\eqref{eq:Lax_eq} is highly non-trivial, and it will be crucial in the first part of this section. 
The first thing that we can think of after the previous section, is to directly compare~\eqref{eq:Lax_eq} with the Hamiltonian equations related to the Lie-Poisson brackets. To achieve such a description, we will need to introduce a second pairing, identified with the brackets~$\langle\,\,\cdot\,\,|\,\,\cdot\,\, \rangle $ establishing the isomorphism $\mathfrak{g}\simeq \mathfrak{g}^*$, following the treatment in~\cite{AdlerBook}. 

We emphasise the difference between the dual pairing $\langle\,\,\cdot\,\,,\,\,\cdot\,\, \rangle $ and the inner product $\langle\,\,\cdot\,\,|\,\,\cdot\,\, \rangle $, and the between $d$ the usual differential, and $\nabla$, the gradient with respect to~$\langle\,\,\cdot\,\,|\,\,\cdot\,\, \rangle $. The natural pairing $\langle\,\,\cdot\,\,,\,\,\cdot\,\, \rangle $ is pairing an element of $\mathfrak{g}^*$ (on the left) with an element of $\mathfrak{g}$ on the right, i.e.
\begin{equation}
    \langle \,L\,,\,X\,\rangle, \qquad X \in \mathfrak{g}, ~~L \in \mathfrak{g}^*\,. 
\end{equation}
Given a function $\varphi(L)\in C^{\infty}(\mathfrak{g}^*)$, $d\varphi(L)$ identifies an element of smooth functions on $\mathfrak{g} \simeq (\mathfrak{g}^*)^*$. In~\eqref{eq:Lie_PB}, we used the differentials built on the natural pairing to introduce the Lie Poisson brackets for functions of elements in $\mathfrak{g}^*$. Whereas, the product~$\langle\,\,\cdot\,\,|\,\,\cdot\,\, \rangle $ is an inner product, i.e.\ it is an ad-invariant ($\text{ad}\simeq \text{ad}^*$) and Ad-invariant ($\text{Ad}\simeq\text{Ad}^*$) non-degenerate symmetric bilinear form, that allows us to identify $\mathfrak{g}^*$ with $\mathfrak{g}$ itself via an isomorphism. For instance, if $\mathfrak{g}$ semi-simple, the inner product can be provided by the Killing form. In general, for $X,Y,Z \in \mathfrak{g}$, we have   
\begin{equation}
    \langle X\,|\,[\,Y,Z\,] \,\rangle = \langle Y\,|\,[\,Z,X\,] \,\rangle = \langle Z\,|\,[\,X,Y\,] \,\rangle \,, \quad \langle \,[\,X,Y\,]\,|\,Z\,\rangle = \langle \,X\,|\,[\,Y,Z\,]\,\rangle \,. 
\end{equation}
This allows us to identify the Lie Poisson bracket~\eqref{eq:Lie_PB} on functions of elements in $\mathfrak{g}^*$ with a Lie bracket on functions of elements in $\mathfrak{g}$, i.e.\ 
\begin{equation}
    \{\,\varphi, \psi \,\}(X) = \langle \,X \,|\, [\,\nabla \varphi(X)\,,\nabla \psi(X)\,]\,\rangle\,, 
\end{equation}
where the symbol $\nabla$ is the gradient of the functions at $X$ with respect to $\langle\,\,\cdot\,\,|\,\,\cdot\,\, \rangle $. The definition of $\nabla$ in terms of $d$ for a function $\varphi \in C^{\infty}(\mathfrak{g})$ is
\begin{equation}
    \langle \, \nabla \varphi(X) \,|\,Y  \,\rangle = \langle \, d\varphi(X)\,, \,Y \, \rangle  \,, \qquad Y \in \mathfrak{g}\,,
\end{equation}
and being $\varphi$ a function on a vector space, this can also be expressed in the form 
\begin{equation}
    \langle \, \nabla \varphi(X) \,|\,Y  \,\rangle = \left. \frac{d}{dt} \varphi(X + t\,Y)  \right|_{t=0}\,. 
\end{equation}
Then, the Hamiltonian vector field on $\mathfrak{g}$ takes a particularly simple form. Indeed, we have on one side, 
\begin{equation}
\begin{split} 
    \{\, \varphi, H\,\}(X) &= \langle\, X \,|\, [\,\nabla \varphi(X),\nabla H(X)\,]\,\rangle \\[1ex]
    &=  \, \langle\, [\, \nabla H(X), \,X  \,]\,| \,\nabla \varphi(X) \rangle = \langle \,d\varphi(X)\,,\,[\,\nabla H(X), X\,]\,\rangle \,, 
\end{split} 
\end{equation}
and from the definition of Hamiltonian vector field, 
\begin{equation}
    \{\, \varphi, H\,\}(X) = \chi_H(X)[\varphi] = \langle \, d\varphi(X) \,,\,\chi_H(X) \, \rangle \,. 
\end{equation}
For the arbitrariness of $\varphi \in C^{\infty}(\mathfrak{g})$, we can equate the right hand sides of the natural pairings, and we find the expression of the Hamiltonian vector field as 
\begin{equation}
    \frac{dX}{dt} = [\,\nabla H(X), X\,]\,,
\end{equation}
which is similar to the form of a Lax equation. 
The sign of the definition depends on the position of $H$ in the Poisson bracket, and it is a mere convention. 

A paradigmatic example of such a construction, comes for the general linear algebra of matrices of order $n$. We consider $\mathfrak{g}=\mathfrak{gl}(n)$ in Proposition~\ref{prop:Ham_eq_Lie_Poisson}, and state the following proposition about trivial Hamiltonian flows. 
\begin{proposition}\label{prop:Lax}
    We consider $\mathfrak{g}=\mathfrak{gl}(n)$ the Lie algebra of $n\times n$ matrices, and the inner product $\langle\,X\,|\,Y\,\rangle = \text{tr}(XY)$ allowing us to identify $\mathfrak{g}^* \simeq \mathfrak{g}$, as in~\eqref{eq:inner_prod}. The dual space is endowed with the Lie-Poisson bracket~\eqref{eq:LiePoisson_bracket} and the Hamiltonian equation of motion for the Hamiltonian function $\varphi \in C^{\infty}(\mathfrak{g})$ is 
    \begin{equation}\label{eq:Casimir}
        \frac{dL}{dt} = [\,\nabla \varphi(L),L\,]\,, 
    \end{equation}
    where $\nabla$ identifies the differential with respect to the inner product $\langle\,\,\cdot\,\,|\,\,\cdot\,\,\rangle$, and the Hamiltonian flow on $\mathfrak{g}$ preserves the spectra of matrices. 
\end{proposition}

Spectral invariants for the matrices are the Casimir functions associated with the Lie-Poisson bracket, and by definition are conserved. Hence, these equations do not describe a dynamics, and also they do not carry any information about integrability of the previous equation for an arbitrary $\varphi$. Also, spectral invariants of the operator $L$ are good candidates to be Hamiltonian functions, but in this case they would give rise to trivial equations of motion. 

\begin{proposition}
    For any Lie algebra $\mathfrak{g}$, a function $\varphi \in C^{\infty}(\mathfrak{g}^*)$ is a Casimir function of the Lie-Poisson bracket on $\mathfrak{g}^*$ if and only if  
    \begin{equation}\label{eq:casimir_coadjoint}
        \text{ad}^*_{d\varphi(L)}L = 0\,, \quad \text{for all }L \in \mathfrak{g}^*\,. 
    \end{equation}
    If $\mathfrak{g}^* \simeq \mathfrak{g}$ via the inner product $\langle\,\,\cdot\,\,|\,\,\cdot\,\,  \rangle$ the expression takes the form $[\,\nabla \varphi(L),L \,]$, i.e.\ the right hand side of~\eqref{eq:Casimir}. 
\end{proposition}
Instead of considering the Lie-Poisson bracket $\{\,\cdot\,\,,\,\,\cdot\,\}$ on $\mathfrak{g}^*$ and its coadjoint orbits as phase spaces, for which the Casimir functions are the spectral invariants, we need to consider a second Lie bracket on $\mathfrak{g}$ and a corresponding Lie-Poisson bracket on $\mathfrak{g}^*$. This second structure is determined in the context of the theory of classical $r$-matrices, and the interplay between the two structures will be fundamental for the description via the Lax equation. 

\subsection{Lie dialgebras} We consider $\mathfrak{g}$ the Lie algebra, and the linear operator~$R \in \text{End}(\mathfrak{g})$, which is called $r$-matrix (or classical $r$-matrix) if the bracket 
\begin{equation}\label{eq:R_matrix}
    [\,X,Y\,]_{R} = 
    \frac{1}{2}
 ([\,R(X),Y\,]+[\,X,R(Y)\,])\,, \qquad X,Y \in \mathfrak{g}\,,
 \end{equation}
is indeed a Lie bracket (skew-symmetric and satisfying the Jacobi identity). 
The equivalent expression in terms of the generic adjoint action over $\mathfrak{g}$ is 
\begin{equation}\label{eq:adjoint_R}
    \text{ad}^R_{X} Y = \frac{1}{2} ( \text{ad}_{R(X)} Y + \text{ad}_{X} R(Y)   )\,.
\end{equation}
The Lie algebra realised through the Lie bracket $[\,\,\cdot\,,\,\cdot\,\,]_R$ is denoted as $\mathfrak{g}_R$. Hence, there are two Poisson brackets living on $\mathfrak{g}^* \simeq \mathfrak{g}^*_R$, $\{\,\cdot\,\,,\,\,\cdot\,\}$ and $\{\,\cdot\,\,,\,\,\cdot\,\}_R$ which are Lie-Poisson brackets on $\mathfrak{g}$ and $\mathfrak{g}_R$ respectively. The pair $(\mathfrak{g},\mathfrak{g}_R)$ is called \emph{dialgebra}\footnote{In older works (also by Semenov-Tian-Shansky) the algebra $\mathfrak{g}_R$ was called \emph{double} of the algebra $\mathfrak{g}$. }. 

A fundamental class of Lie dialgebras is given if $\mathfrak{g}$ admits a vector space decomposition in the sum of two subalgebras
\begin{equation}
    \mathfrak{g}=\mathfrak{g}_+ \oplus \mathfrak{g}_-\,.  
\end{equation}
With $P_{\pm}$ the projection operator onto $\mathfrak{g}_{\pm}$ along the complementary subalgebra the linear operator $R$ is 
\begin{equation}\label{eq:R_pm_I_pm}
    R= P_+ - P_- \,, \qquad I = P_+ + P_-\,,
\end{equation}
and $I$ the identity. 
From~\eqref{eq:R_matrix} the Lie bracket $[\,\,\cdot\,,\,\cdot\,\,]_R$ is 
\begin{equation}
    [\,X,Y\,]_R = [\,X_+,Y_+\,]-[\,X_-,Y_-\,]\,, \qquad Z_{\pm} = P_{\pm} Z\,,~ Z \in \mathfrak{g}\,, 
\end{equation}
and in terms of a generic adjoint action on $\mathfrak{g}$ 
\begin{equation}\label{eq:adjoint_R_pm}
    \text{ad}^R_X Y = \text{ad}_{X_+} Y_+ - \text{ad}_{X_-} Y_- \,, \qquad X_{\pm}, Y_{\pm} \in \mathfrak{g}_{\pm}\,.  
\end{equation}
The motivation to introduce the notion of the Lie dialgebra is found in the following fundamental theorem.
\begin{theorem}\label{thm:involutivity_R}
    Let $I(\mathfrak{g})$ be the ring of Casimir functions on $\mathfrak{g}^*$. We have: 
    \begin{enumerate}[label=\textup{($\roman*$)}]
        \item any two functions $\varphi,\psi \in I(\mathfrak{g})$ are in involution with respect to $\{\,\cdot\,\,,\,\,\cdot\,\}_R$ on $\mathfrak{g}^*$; 
        \item the equation of motion induced by a function $\varphi \in I(\mathfrak{g})$ with respect to $[\,\cdot\,\,,\,\,\cdot\,]_R$ are 
        \begin{equation}
            \frac{dL}{dt} = - \textup{ad}^*_{M} L\,, \qquad M(L) = \frac{1}{2} R(d\varphi(L));
        \end{equation}
        \item if $\mathfrak{g}$ admits the inner product $\langle\,\cdot\,,\,\cdot\, \rangle$ such that $\mathfrak{g} \simeq \mathfrak{g}^*$ we have instead 
        \begin{equation}
            \frac{dL}{dt} = [\, L,M  \,]\,, \qquad M(L)= \frac{1}{2} R(\nabla \varphi(L))\,.
        \end{equation}
    \end{enumerate}
\end{theorem}
We follow the definition for Lie-Poisson bracket~\eqref{eq:Lie_PB} to construct $\{\,\cdot\,\,,\,\,\cdot\,\}_R$ for functions on $\mathfrak{g}^*$, i.e.\ 
\begin{equation}\label{eq:LiePoisson_R}
\begin{split}
    \{ \varphi , \psi\}_R (L) &= \langle L , \text{ad}^R_{d\varphi(L)}  d\psi(L) \rangle  \,,
\end{split}
\end{equation}
with the adjoint action expressed in~\eqref{eq:adjoint_R}. To show $(i)$, by definition a function $\varphi \in I(\mathfrak{g})$  is such that~\eqref{eq:casimir_coadjoint}. We consider an arbitrary $X \in \mathfrak{g}$ and have 
\begin{equation}
\begin{split} 
    0=(\text{ad}^*_{d\varphi(L)} L)(X) &= \langle \text{ad}^*_{d\varphi(L)} L , X\rangle =- \langle L ,  \text{ad}_{d\varphi(L)} X\rangle \,.
\end{split} 
\end{equation}
Therefore $\text{ad}_{d\varphi(L)} X= 0$ for all $X \in \mathfrak{g}$, and making~\eqref{eq:LiePoisson_R} explicit, we have the proof. 

To show $(ii)$ we consider a function $\varphi \in C^{\infty}(\mathfrak{g}^*)$ and the Hamiltonian $H \in I(\mathfrak{g})$ with Hamiltonian vector field defined with respect to $\{\,\cdot\,\,,\,\,\cdot\,\}_R$ such that the equation of motion for $\varphi$ is 
\begin{equation}
    \frac{d\varphi}{dt}(L) = \{H, \varphi \}_R(L) = \langle L , \text{ad}^R_{dH(L)}  d\varphi(L) \rangle  \,.
\end{equation}
Making the adjoint action of $\mathfrak{g}_R$ explicit as in~\eqref{eq:adjoint_R}, we have 
\begin{equation}
    \begin{split}
        \frac{d\varphi}{dt}(L) &= \langle L , \text{ad}^R_{dH(L)}  d\varphi(L) \rangle \\
        &= \frac{1}{2} (\langle L , \text{ad}_{R(dH(L))}  d\varphi(L) \rangle + \langle L , \text{ad}_{dH(L)}  R(d\varphi(L)) \rangle) \\
        &= \frac{1}{2} \langle L , \text{ad}_{R(dH(L))}  d\varphi(L) \rangle = - \frac{1}{2} \langle \text{ad}^*_{R(dH(L))} L ,   d\varphi(L) \rangle  \,.
    \end{split}
\end{equation}
By taking $\varphi$ to be a set of linear coordinates in $\mathfrak{g}^*$, we have the equation of motion of $L$ 
\begin{equation}\label{eq:eq_motion_L}
    \frac{dL}{dt} = - \frac{1}{2} \text{ad}^*_{R(dH(L))} L\,. 
\end{equation}
In the case where the inner product $\langle\,\cdot\,,\,\cdot\,\rangle$ is defined, for $(iii)$ we get $\text{ad}^* = \text{ad}$ and the equation of motion for $L \in \mathfrak{g}$ becomes 
\begin{equation}
    \frac{dL}{dt} = [\,L, R(\nabla H(L))\,]\,, 
\end{equation}
recalling that $\nabla$ is the differential sign with respect to the bilinear form $\langle\,\cdot\,|\,\cdot\,\rangle$, analogous to the sign $d$ which represents the differential with respect to the natural pairing $\langle\,\cdot\,,\,\cdot\,\rangle$, as introduced in Proposition~\ref{prop:Lax}.   

From the expression of the operator $R$ and the identity in terms of the projections $P_{\pm}$ as in~\eqref{eq:R_pm_I_pm} we get 
\begin{equation}
    R(dH(L)) = (P_+-P_-)(dH(L)) = (I \pm 2 P_{\pm})(dH(L))\,, 
\end{equation}
and then using again the fact that $H \in I(\mathfrak{g})$, we get 
\begin{equation}\label{eq:eq_motion_L_pm}
     \frac{dL}{dt} = \mp \,\text{ad}^*_{(dH(L))_{\pm}}  L=- \text{ad}^*_{M_{\pm}(L)}L \,, \quad M_{\pm}(L) = \frac{1}{2}(R\pm I)(dH(L))\,. 
\end{equation}

From the geometric point of view, the trajectories of the dynamical system with Hamiltonian $H \in I(\mathfrak{g})$ are in the intersection of two families of orbits in $\mathfrak{g}^*$: the coadjoint orbits of $\mathfrak{g}$ and $\mathfrak{g}_R$. In many cases, the intersections of orbits are exactly the Liouville tori for the dynamical system.  

\subsection{Factorisation theorem}
One of the crucial features of the inverse scattering method is the reduction of the equations of motion to the Riemann problem. This is already somehow underpinned in the approach we considered, indeed a version of the Riemann problem is provided by the factorisation problem at the level of Lie groups. We can then define a global version of Theorem~\ref{thm:involutivity_R} by introducing $G$ the Lie group with Lie algebra $\mathfrak{g}$, and $G_{\pm}$ the subgroups associated with the subalgebras~$\mathfrak{g}_{\pm}$. 

\begin{theorem}
    Let $H \in I(\mathfrak{g})$, $X=dH(L)$. We denote with $g_{\pm}(t)$ the smooth curves in $G_{\pm}$ that are solutions of the factorisation problem 
    \begin{equation}\label{eq:factor_group}
        \exp(tX) = g_+(t)\, g_-(t)^{-1}\,, \qquad g_{\pm}(0) = e \,.
    \end{equation}
    Then, the integral curve $L(t)$ of the equation of motion~\eqref{eq:eq_motion_L_pm} with initial condition $L(0)=\Lambda$ is 
    \begin{equation}\label{eq:Adjoint_L}
        L(t) = \textup{Ad}^*_{g_{\pm}(t)^{-1}} \Lambda \,. 
    \end{equation}    
\end{theorem}

We consider the expression in~\eqref{eq:Adjoint_L} on an arbitrary element $X \in \mathfrak{g}$
\begin{equation}
    L(t)(X) = \langle \text{Ad}^*_{g_{\pm}(t)^{-1}}\Lambda\,, X \rangle  = \langle \Lambda\,, \text{Ad}_{g_{\pm}(t)}X \rangle  \,,
\end{equation}
and we differentiate this expression with respect to the parameter $t$. In order to make the computation more transparent for the intermediate steps we will assume that the subgroups $G_{\pm}$ are matrix groups, so that the adjoint action of $G_{\pm}$ on the Lie algebra $X$ can be expressed in terms of the conjugation: 
\begin{equation*}
    \begin{split}
        \frac{dL(t)}{dt}(X) &= \frac{d}{dt}\langle \Lambda\,, \text{Ad}_{g_{\pm}(t)}X \rangle = \frac{d}{dt}\langle \Lambda\,, {g_{\pm}(t)}X {g_{\pm}(t)}^{-1} \rangle \\
        &= \langle \Lambda\,, {\dot{g}_{\pm}(t)}X {g_{\pm}(t)}^{-1} + {{g}_{\pm}(t)}X {\dot{g}_{\pm}(t)}^{-1}\rangle \\
        &= \langle \Lambda\,, {g_{\pm}(t)}{g_{\pm}(t)}^{-1}\,{\dot{g}_{\pm}(t)}\,X {g_{\pm}(t)}^{-1} + {{g}_{\pm}(t)}X \,{\dot{g}_{\pm}(t)}^{-1} {g_{\pm}(t)} {g_{\pm}(t)}^{-1}   \rangle \,.
    \end{split}
\end{equation*}
By using the fact that $g\,\dot{g}^{-1}=-\dot{g}\,g^{-1}$ we get 
\begin{equation}
    \begin{split}
         \frac{dL(t)}{dt}(X) &= \langle \Lambda\,, {g_{\pm}(t)}\,[\, {{g}_{\pm}(t)}^{-1}\,{\dot{g}_{\pm}(t)},X  \,]\,{g_{\pm}(t)}^{-1}\rangle \\
         &=  \langle \Lambda\,, \text{Ad}_{g_{\pm}(t)}( \text{ad}_{g_{\pm}(t)^{-1}\,\dot{g}_{\pm}(t)}  X) \rangle \\
         &= \langle \text{Ad}^*_{g_{\pm}(t)^{-1}} \Lambda\,, \text{ad}_{g_{\pm}(t)^{-1}\,\dot{g}_{\pm}(t)}  X\rangle  \\
         &= \langle L\,, \text{ad}_{g_{\pm}(t)^{-1}\,\dot{g}_{\pm}(t)} X\rangle =  -\langle \text{ad}^*_{g_{\pm}(t)^{-1}\,\dot{g}_{\pm}(t)} L\,,  X \rangle\,,
    \end{split}
\end{equation}
and from the arbitrariness of $X \in \mathfrak{g}^*$ we get 
\begin{equation}
    \frac{dL}{dt} = - \text{ad}^*_{g_{\pm}(t)^{-1}\,\dot{g}_{\pm}(t)} L \,,
\end{equation}
and for matrix groups with inner product we have 
\begin{equation}
    \frac{dL}{dt} = [\,L , g_{\pm}(t)^{-1}\,\dot{g}_{\pm}(t)\,]\,.
\end{equation}
We have to still check that $M_{\pm}=\pm P_{\pm}X(t)$ in~\eqref{eq:eq_motion_L_pm} is indeed $g_{\pm}(t)^{-1}\,\dot{g}_{\pm}(t)$. We consider the factorisation problem
\begin{equation}
    g_+(t)\,g_-(t)^{-1} = \exp(tX) \,,~~ \implies ~~ g_+(t)= \exp(tX)\,g_-(t)\,, 
\end{equation}
and we derive the expression with respect to the parameter $t$, yielding 
\begin{equation}
    \text{Ad}_{g_{+}(t)}^{-1} X =  g_{+}(t)^{-1}\,\dot{g}_{+}(t) - g_{-}(t)^{-1}\,\dot{g}_{-}(t)\,. 
\end{equation}
Begin $X = dH(t)$ an $\text{Ad}~ \mathfrak{g}$-invariant function, we have 
\begin{equation}
    X(t) = \text{Ad}_{g_{\pm}(t)^{-1}} X \,,
\end{equation}
and this adds up to 
\begin{equation}
    X(t) =  g_{+}(t)^{-1}\,\dot{g}_{+}(t) - g_{-}(t)^{-1}\,\dot{g}_{-}(t) \,,
\end{equation}
so that $\pm P_{\pm} X(t) = g_{\pm}(t)^{-1}\,\dot{g}_{\pm}(t)$.

We now consider the factorisation theorem in the Hamiltonian reduction, that allows us to focus more on the geometric aspects of the approach. 
This description requires three steps: 
\begin{enumerate}
    \item choose a larger phase space who would generate the Lax system as a reduced symplectic manifold;
    \item identify a symmetry group and the free dynamics possessing the suitable invariance;
    \item describe the reduced system.
\end{enumerate}
The natural choice of the large phase space is the cotangent bundle $T^*G$, and we consider the free dynamics on it. As we did in the previous section, we fix the left trivialisation $T^*G \simeq G \times \mathfrak{g}^*$. 

\begin{lemma}
    A Casimir function $H \in I(\mathfrak{g})$ admits a canonical lift to $T^*G$ by the following 
    \begin{equation}
        \chi_H = H \circ \mu_{\lambda} \,, \qquad \chi_H = H \circ \mu_{\rho} \,,
    \end{equation}
with $\mu_{\lambda}$, $\mu_{\rho}$ are the moments in~\eqref{eq:moments}. The resulting functions $\chi_H$ may be characterised as bi-invariant functions on $T^*G$. 
\end{lemma}

\begin{lemma}
    In left trivialisation $T^*\simeq G \times \mathfrak{g}^*$, the Hamiltonian flow $F_t$ generated by a bi-invariant Hamiltonian $\chi_H$ is given by 
    \begin{equation}\label{eq:flow}
        F_t \colon (g,L) \mapsto (g\, \exp(t\,dH(L)),L) \,, ~ g \in G, ~ L \in \mathfrak{g}^*, 
    \end{equation}
    i.e.\ the integral curves projects onto the left translates of on-parameter subgroup in $G$. The choice of $H$ determines the dependence of the (constant) velocity vector $dH(L)$ on the initial canonical momentum $L$. 
\end{lemma}
Let us consider the dual pair 
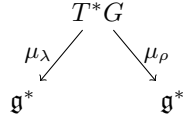
\begin{equation*}
    \begin{tikzpicture}
              \path (0,0) node (a) {$T^*G$} -- ++(-1,-1.2) node (b) {$\mathfrak{g}^*$} -- ++(2,0) node (c) {$\mathfrak{g}^*$};
            \draw[->] (a) -- (b) node[midway,left] {\small $\mu_{\lambda}$}; 
            \draw[->] (a) -- (c) node[midway,right] {\small $\mu_{\rho}$}; 
    \end{tikzpicture}
\end{equation*}
The Hamiltonian $\chi_H$ is constant by construction on the fibres of the projection maps $\mu_{\lambda}$ and $\mu_{\rho}$. For both maps the reduced Hamiltonians coincide with $H$, hence they are Casimir functions of the quotient Poisson structure. 

Since the Hamiltonian $\chi_H$ is invariant with respect to the action of $G$ by left and right translations, the flow $F_t$ admits reductions with respect to any subgroup $U \in G \times G$. Different subgroups give rise to different reduced systems. The particular choice leading to the Lax system~\eqref{eq:eq_motion_L_pm} corresponds to the choice of the $r$-matrix, i.e.\ the Lie group associated with the Lie algebra $\mathfrak{g}_R$ 
\begin{equation}
    U = G_+ \times G_- \simeq G_R \,. 
\end{equation}
The action of $G_R$ on $T^*G$ based on the expression of left and right translations in left trivialisation~\eqref{eq:lambda_rho} is 
\begin{equation}\label{eq:action_G_R}
    (h_+,h_-) \colon (g,L) \mapsto (h_+\,g\,h^{-1}_-, \text{Ad}^*_h \, L)\,.
\end{equation}
To realise a reduction over $G_{R}$ we construct a map $s$ constant on the orbits of $G_R$ in $T^*G$, so that its image will provide a model of the quotient space. 

We consider $g \in G$ and $g_{\pm}\in G_{\pm}$ the solutions of the factorisation problem
\begin{equation}
    g = g_+ \, g_-^{-1}\,. 
\end{equation}
Then we have the following theorem, collecting the previous results. 
\begin{theorem} In left trivialisation we consider the group $G_R \simeq G_+ \times G_-$. We have: 
\begin{enumerate}[label=\textup{($\roman*$)}]
    \item the Hamiltonian $\chi_H$ is invariant with respect to the action~\eqref{eq:action_G_R};
    \item the map $s \colon T^*G \to \mathfrak{g}^*$ that works as 
    \begin{equation*}
        (g,L) \mapsto \textup{Ad}^*_{g_-^{-1}} \,L 
    \end{equation*}
    is constant on $G_R$-orbits in $T^*G$;
    \item the quotient Poisson structure on $T^*G/G \simeq \mathfrak{g}^*$ coincide exactly with the Lie-Poisson bracket $\{\,\cdot\,,\,\cdot\,\}_{R}$ of $\mathfrak{g}_R$; 
    \item the quotient flow $\bar{F}_t$ on ${\mathfrak{g}^*_R}$ obtained by the free flow $F_t$ in \eqref{eq:flow} as 
    \begin{equation*}
        \bar{F}_t \colon L \mapsto \textup{Ad}^*_{g_{\pm}(t)^{-1}} L\,,
    \end{equation*}
    with $g_{\pm}(t)$ solutions of the factorisation problem 
    \begin{equation}\label{eq:factor_subgroups}
        \exp(t\, dH(L)) = g_+(t)g_{-}(t)^{-1}
    \end{equation}
    and satisfy the Lax equation~\eqref{eq:eq_motion_L_pm};
    \item the reduced Hamiltonian on $\mathfrak{g}^*_R$ coincides with $H$. 
\end{enumerate}
\end{theorem}

This theorem implies that the Lax equations obtained by reduction of bi-invariant Hamiltonian equations on the cotangent bundle $T^*G$ are explicitly solvable. The question whether these are completely integrable in the Liouville sense needs to be addressed separately. The Liouville-Arnold theorem~\cite{Arnold} establishes that a Hamiltonian system on a $2n$-symplectic manifold is completely integrable if it admits exactly $n$ independent first integrals of motion. In this case, the generic level surface of these integrals is a \emph{Lagrangian submanifold}. 

Let $(\mathcal{M},\omega)$ be a $2n$-dimensional symplectic manifold with the integrals of motion Casimir functions on $\mathfrak{g}^*$, say $f_1=H$, $f_2,\dots,f_n$. Let $c \in \mathbb{R}^n$ be a regular value of $f:=(f_1,\dots,f_n)$. Then the corresponding level $f^{-1}(c)$ is a Lagrangian submanifold of $\mathcal{M}$. This means that locally around the regular value $c$, the map $f\colon \mathcal{M}\to \mathbb{R}^n$ collecting the integrals of motion is a \emph{Lagrangian fibration}, i.e.\ it is locally trivial and the fibres are Lagrangian submanifolds. Also, one can show that the connected components of $f^{-1}(c)$ are of the form $\mathbb{R}^{n-k}\times \mathbb{T}^k$, where $0 \leq k \leq n$ and $\mathbb{T}^k$ is a $k$-dimensional torus. In particular, every compact component must be a Lagrangian torus. 

In our description, if $G$ is a finite-dimensional semi-simple group, the ring of its coadjoint invariants $I(\mathfrak{g})$ is finitely generated, and the Chevalley theorem establishes that the number of independent generators is $\ell = \text{rank}G$. In the case of $G_R$ though the number of generators is $\ell^2$, and hence the number of functions in involution derived in Theorem~\ref{thm:involutivity_R} is not sufficient. Nevertheless, usually the Lax equations are much more regular and the associated Hamiltonian systems in the case of a finite-dimensional semi-simple group $G$ are \emph{degenerate integrable}. To complete the set of integrals in involution, one can consider several sets of so-called \emph{semi-invariants}. This difficulty does not arise in the case of Lax systems on low dimensional coadjoint orbits of the bracket $\{\,\cdot\,,\,\cdot\,\}_R$ (i.e.\ open Toda lattices). 

\subsection{\boldmath \texorpdfstring{Classical Yang-Baxter equations and $r$-matrices (I)}{YB}}

In this section, we will deepen one of the fundamental characteristics of the $r$-matrices, i.e.\ begin solutions to the modified classical Yang-Baxter equation. 

We consider the Lie bracket $[\,\cdot\,,\,\cdot\,]_R$ in~\eqref{eq:R_matrix}. The Jacobi identity for this bracket is given by 
\begin{equation}\label{eq:Jacobi_for_R}
    [\bar{R}(X,Y),Z] + [\bar{R}(Z,X),Y] + [\bar{R}(Y,Z),X] = 0 \,,  
\end{equation}
with $\bar{R}\colon \mathfrak{g} \to \mathfrak{g}$ the  skew-symmetric bilinear form defined as 
\begin{equation}
\begin{split}
    \bar{R}(X,Y) &= [R(X),R(Y)] - 2R([X,Y]_R)  \\
    &= [R(X),R(Y)] - R([R(X),Y]+[X,R(Y)])  \,. 
\end{split}
\end{equation}
Usually the trilinear condition expressed in~\eqref{eq:Jacobi_for_R} is substituted by a bilinear sufficient condition in terms of the so-called classical Yang-Baxter equation (CYBE)
\begin{equation}\label{eq:CYBE}
 \begin{split} 
    \bar{R}(X,Y) &= 0 \,, \\
    [R(X),R(Y)]&=R([R(X),Y]+[X,R(Y)])\,.
  \end{split} 
\end{equation}
Another important sufficient condition is given by the modified classical Yang-Baxter equation (mCYBE) 
\begin{equation}\label{eq:mCYBE}
	\begin{split} 
     \bar{R}(X,Y) &= -c[X,Y]\,,\\
     [R(X),R(Y)]-R([R(X),Y]+[X,R(Y)])&=-c[X,Y]\,.
     \end{split} 
\end{equation}
In particular, if $\mathfrak{g}$ is a Lie algebra on real values $c=\pm 1$. The $r$-matrix $R$ presented in the previous section satisfies~\eqref{eq:mCYBE} with $c= 1$, that is also called the \emph{split case}. The $r$-matrices satisfying~\eqref{eq:mCYBE} not always admit the simple formulation as in ~\eqref{eq:R_pm_I_pm}, but they are still associated to a factorisation problem. On the contrary, the case~\eqref{eq:CYBE} corresponds to a degenerate case, and is not associated with a degeneration problem. The case $c=-1$ is called \emph{non-split case}, a factorisation problem is still admissible, but with some different specification for the factors. 

Let us consider the factorisation problem associated to a $r$-matrix $R \in \text{End}(\mathfrak{g})$ satisfying the mCYBE (split case). We can define then 
\begin{equation}
    R_{\pm} = \frac{1}{2}(R \pm I) \,, 
\end{equation}
so that we can show that  
\begin{equation}
    [R_{\pm}(X),R_{\pm}(Y)] = R_{\pm}[X,Y]\,, \qquad X,Y \in \mathfrak{g}\,, 
\end{equation}
i.e.\ $R_{\pm} \colon \mathfrak{g}_R \to \mathfrak{g}$ is a homomorphism. Let us set 
\begin{equation}
    \mathfrak{g}_{\pm} = \text{Im} R_{\pm} \,, \qquad \mathfrak{t}_{\pm} = \text{Ker} R_{\mp}\,. 
\end{equation}
Then we have that $\mathfrak{g}_{\pm}$ is a subalgebra and $\mathfrak{t}_{\pm}$ is an ideal. We introduce the map $\theta_R \colon \mathfrak{g}_+/\mathfrak{t}_+ \to \mathfrak{g}_-\mathfrak{t}_- $ acting as: 
\begin{equation}
    \theta_R \colon (R + I)X \mapsto (R-I)X\,, \qquad X \in \mathfrak{g}\,. 
\end{equation}
\begin{proposition}
    The map $\theta_R$ is a Lie algebra isomorphism. The combined map $i_R = R_+ \oplus R_-$ defined as 
    \begin{equation}
        i_R \colon \mathfrak{g}_R \to \mathfrak{g} \oplus \mathfrak{g} \,, \qquad 
        X \mapsto (R_+(X), R_-(Y)) 
    \end{equation}
    is a Lie algebra embedding, and $\text{Im}(i_R)$ is a Lie subalgebra of $\mathfrak{g}_+\oplus  \mathfrak{g}_-$. The composition of the map $i_R$ and the map $\alpha$ defined as 
    \begin{equation}
        \alpha \colon \mathfrak{g} \oplus \mathfrak{g} \to \mathfrak{g}_+ \oplus \mathfrak{g}_-\,, \qquad (X_+, X_-) \mapsto X_+ - X_-\,, 
    \end{equation}
    provides a unique decomposition of any element $X \in \mathfrak{g}$ as $X = R_+(X)-R_-(X)$. 
\end{proposition}
The previous is based on the observation that $I = R_+ - R_-$ is the identity mapping. 

The homomorphism $R_{\pm}$ on the algebras $\mathfrak{g}$ and $\mathfrak{g}_R$ give rise to Lie group homomorphisms on the corresponding Lie groups $G$, $G_R$. We introduce 
\begin{equation}
    G_{\pm} = R_{\pm}(G) \,, \qquad K_{\pm} = \text{Ker} R_{\mp}(G)\,, 
\end{equation}
and extend the map $\theta_R$ to a Lie group isomorphism $\theta_R \colon G_+/K_+ \to G_-/K_-$\,. Then we construct the map 
\begin{equation}
    i_R \colon G_R \to G \times G \,, \qquad h \mapsto (R_+\, h, R_-\,h)\,, 
\end{equation}
which is a Lie group embedding. We consider the map $\alpha$ defined as 
\begin{equation}
        \alpha \colon G_+ \times G_- \to G \,, \qquad (x,y) \mapsto x\,y^{-1}, 
\end{equation}
and again the composition of the maps $i_R$ and $\alpha$ is a local homomorphism, and an element $x \in G$ admits a unique decomposition 
\begin{equation}\label{eq:fact}
    x = x_+\, x_-^{-1} \,, \qquad (x_+, x_-) \in \text{Im} (R_+ \times R_-)\,. 
\end{equation}

\subsection{Double of Lie dialgebras}
Usually solutions to the mCYBE are not expressible in the simple form~\eqref{eq:R_pm_I_pm} as the difference of two complementary projection operators. It is still possible to reduce the problem to the case~\eqref{eq:R_pm_I_pm} by squaring the initial Lie algebra. Let $R \in \text{End}(\mathfrak{g})$ be a solution of the mCYBE. We set $\mathfrak{d}= \mathfrak{g} \oplus \mathfrak{g}$ and from the previous section, the Lie algebra $\mathfrak{g}_R$ is canonically embedded in $\mathfrak{d}$. Let $\mathfrak{g}^{\delta}$ be the diagonal subalgebra. We have the following proposition. 

\begin{proposition}
    The algebra $\mathfrak{d}$ admits the decomposition 
    \begin{equation}
        \mathfrak{d} = \mathfrak{g}^{\delta} \oplus \mathfrak{g}_R\,.
    \end{equation}
    Conversely, any Lie subalgebra $\mathfrak{h}\subset \mathfrak{g} \oplus \mathfrak{g}$ which is transversal to the diagonal gives rise to a solution of the mCYBE. 
\end{proposition}
We can derive the expression for the projection operators $P_{\mathfrak{g}^{\delta}}$ and $P_{\mathfrak{g}_R}$, i.e.\ for $(X,Y) \in \mathfrak{d}$
\begin{equation}
    \begin{split}
        P_{\mathfrak{g}_R} &= (X_+ - Y_+, X_- - Y_-) \,, \\
        P_{\mathfrak{g}^{\delta}} &= (Y_+ - X_-, Y_+ - X_-) \,, 
    \end{split}
\end{equation}
with $X_{\pm}=R_{\pm}(X)$, $Y_{\pm}=R_{\pm}(Y)$. We introduce the canonical projections $P_{\pm}\colon \mathfrak{g} \oplus \mathfrak{g} \to \mathfrak{g}$ selecting the first and second component of $(X,Y)\in \mathfrak{d}$ respectively. We set $R_{\pm} = \left.  P_{\pm} \right|_{\mathfrak{h}}$, so that $R_{\pm} \colon \mathfrak{h} \to \mathfrak{g}$ is a Lie algebra homomorphism and $R_+ - R_-$ defines an isomorphism between $\mathfrak{h}$ and $\mathfrak{g}$. Hence, the pair $(\mathfrak{g},\mathfrak{h})$ is a Lie dialgebra and $R=(R_++R_-)/2$ satisfies mCYBE. 

We introduce 
\begin{equation}
    r_{\mathfrak{d}} = P_{\mathfrak{g}_R} - P_{\mathfrak{g}^{\delta}}\,, 
\end{equation}
which satisfies mCYBE and endows $\mathfrak{d}$ with a Lie dialgebra structure expressed as the pair $(\mathfrak{d},\mathfrak{d}_{R_{\mathfrak{d}}})$ which is called the \emph{double} of $(\mathfrak{g},\mathfrak{g}_R)$. 

Let us introduce the notation for the corresponding Lie groups associated with the Lie algebras $\mathfrak{d}$ and $\mathfrak{g}^{\delta}$, i.e.\ $D = G \times G$ and $G^{\delta} \subset D$. Then, an arbitrary element $(x,y) \in D$  sufficiently close to the unity admits a unique factorisation 
\begin{equation}
    (x,y) = (h_+,h_-)(g,g)\,, \qquad (h_+,h_-) \in G_R\,,~ g \in G\,, 
\end{equation}
where 
\begin{equation}\label{eq:hpm}
    h_{\pm} = (x\,y^{-1})_{\pm} \,, \qquad g= x(x\,y^{-1})_+^{-1} = y(x\,y^{-1})_-^{-1}\,.
\end{equation}

Now we finally describe the Lax equations on $\mathfrak{d}^*$ starting from their expression on $\mathfrak{g}^*$. We have that the dual Lie algebra is 
\begin{equation}
    \mathfrak{d}^* = (\mathfrak{g}^{\delta})^* \oplus \mathfrak{g}_R^* \,,  
\end{equation}
and by construction we can immediately see that 
\begin{equation}
    (\mathfrak{g}^{\delta})^* = \mathfrak{g}_R^{\perp}\,, \qquad \mathfrak{g}_R^* = (\mathfrak{g}^{\delta})^{\perp}\,. 
\end{equation}

\begin{proposition}
    Let $\varphi \in I(\mathfrak{d})$ be a Casimir function of $\mathfrak{d}$. The generalised Lax equation associated with $\varphi$ with respect to $\{\,\cdot\,,\,\cdot\,\}_{R_{\mathfrak{d}}}$ leaves the subspace $\mathfrak{g}_R^* \subset \mathfrak{d}^*$ invariant and the induced vector field on $\mathfrak{g}_R^*$ is Hamiltonian with respect to $\{\,\cdot\,,\,\cdot\,\}_{R}$ with the Hamiltonian $\left. \varphi\right|_{\mathfrak{g}_R^*}$. 
\end{proposition}
We make use of the pairing between $\mathfrak{d}$ and $\mathfrak{d}^*$ expressed as 
\begin{equation}
    \langle (\xi,\eta),(X,Y) \rangle = \langle \xi,X \rangle - \langle \eta,Y \rangle \,, \qquad (X,Y) \in \mathfrak{d},~ (\xi,\eta) \in \mathfrak{d}^*\,, 
\end{equation}
and in this realisation we can determine the complements of $\mathfrak{g}^{\delta}$ and $\mathfrak{g}_R$\, i.e\ 
\begin{equation}
\begin{split}
    (\mathfrak{g}^{\delta})^* &= \{ (\xi,\xi) , \xi \in \mathfrak{g}^* \}\,, \\
    \mathfrak{g}_R^* &= \{ (R^*_-\xi,R^*_+\xi) , \xi \in \mathfrak{g}^* \} \,. 
\end{split}
\end{equation}
Hence we have $(\xi,\xi) \in \mathfrak{g}_R^* \simeq (\mathfrak{g}^{\delta})^{\perp}$. If we set $d\varphi(\xi,\xi)=(X,Y)$ we have that the restriction to $\mathfrak{g}_R^*$ is given by 
\begin{equation}
    \left. d\varphi \right|_{\mathfrak{g}_R^*}(\xi) = X - Y \,, 
\end{equation}
and from the definition of a Casimir function 
\begin{equation}
    \text{ad}^*_{d\varphi_{(\xi,\xi)}} (\xi,\xi) = 0 ~~ \implies ~~ \text{ad}^*_{X-Y} \xi = 0  \,, 
\end{equation}
so $\varphi$ is a Casimir function for $\mathfrak{g}_R^*$ as well. To see that Lax equations generated by $\varphi$ and $\left.\varphi\right|_{\mathfrak{g}^*_R}$ coincide on $\mathfrak{g}_R^* \subset \mathfrak{d}^*$, we compare the factorisation problems in $D$ and $G$. The factorisation on $D$ gives 
\begin{equation}
    (h_{+}(t),h_{-}(t))\,(g(t),g(t)) = \exp(t\, d\varphi(\xi,\xi))\,, \quad (h_{+},h_{-}) \in G_{R}\,,~(g,g) \in G^{\delta}\,, 
\end{equation}
so that the integrable curve $(\xi(t),\xi(t))$ emerging from $(\xi,\xi)$ is 
\begin{equation}
    (\xi(t),\xi(t)) = \text{Ad}^*_{(h_+(t),h_-(t))^{-1}} (\xi,\xi) = (\text{Ad}^*_{h_+(t)^{-1}} \xi\,,\,\text{Ad}^*_{h_-(t)^{-1}} \xi)\,, 
\end{equation}
and from~\eqref{eq:hpm} we have 
\begin{equation}
    h_{\pm}(t) = \exp(t\, d\varphi(\xi,\xi))_{\pm} = \exp(t\, (X-Y)))_{\pm} = \exp(t\, d\left.\varphi\right|_{\mathfrak{g}^*_R})_{\pm} \,. 
\end{equation}
One can properly express the coadjoint representation of $G_R$ in terms of the action of $G$ and the action on $\mathfrak{g}^*_R$ in terms of the action on $\mathfrak{g}^*$ , i.e.
\begin{equation}\label{eq:coadj_R}
\begin{split}
    \text{Ad}^{R*}_{(h_+,h_-)}\xi &= R^*_+(\text{Ad}^*_{h_+}\xi) - R^*_-(\text{Ad}^*_{h_-}\xi)\,, \qquad (h_+,h_-)\in G_R\,,\\
    \text{ad}^{R*}_{X}\xi &= R^*_+(\text{ad}^*_{X_+}\xi) - R^*_-(\text{ad}^*_{X_-}\xi)\,, \qquad X = X_+-X_- \in \mathfrak{g}^*_R\,. 
\end{split}
\end{equation}
\begin{exercise}
	Derive the expression of the coadjoint representation of $G_R$ in~\eqref{eq:coadj_R}.
\end{exercise}
Finally, the Lax equations can be expressed as 
\begin{equation}\label{eq:lax_R}
    \frac{dL}{dt} = - \text{ad}^{R*}_{d\varphi(L)} L = - \text{ad}^{*}_{R_{\pm}(d\varphi(L))} L \,, \qquad L \in \mathfrak{g}^*\,. 
\end{equation}
The presented approach of the double of a Lie dialgebra can be seen as a version of a similar construction introduced by Drinfeld for Lie bialgebras.  
\begin{exercise}
	Let $\sigma \colon G_R \to G $ be the local diffeomorphism defined by the factorization problem~\eqref{eq:fact}. The solution of the Lax equation~\eqref{eq:lax_R} is given by
	\begin{equation}
		L(t)=\text{Ad}^*_{G_R} \sigma^{-1}(\text{exp} (t\,d\varphi(L_0))) L_0\,. 
	\end{equation}
\end{exercise}

\section{\boldmath \texorpdfstring{$r$-matrices for Lie bialgebras}{rMatbi}}\label{sec:r_mat_bialgebra}

The definition of bialgebras and dialgebras are different and related to different notions of classical $r$-matrices as we will see. In the case of Lie dialgebras $(\mathfrak{g},\mathfrak{g}_R)$ we have two Lie algebra structures $[\,\cdot\,,\,\cdot\,]$ and $[\,\cdot\,,\,\cdot\,]_R$ on the same linear space, and the associated $r$-matrix $R \in \text{End}(\mathfrak{g})$.  In the case of Lie bialgebras $(\mathfrak{g},\mathfrak{g}^*)$ the brackets $[\,\cdot\,,\,\cdot\,]$ and $[\,\cdot\,,\,\cdot\,]_*$ are defined on dual linear spaces and $r \in \mathfrak{g} \otimes \mathfrak{g}$. Nevertheless, these two structures can be deeply related, as we will see in the last part of this section by using the language of maps. Here, we can already state the following proposition.  

\begin{proposition}
	Let $(\mathfrak{g},\mathfrak{g}_R)$ be a Lie dialgebra, with $\mathfrak{g}$ endowed with an invariant inner product $\langle \,\cdot\,,\,\cdot\,\rangle$ allowing the identification $\mathfrak{g}^* \simeq \mathfrak{g}$, and $R \in \text{End}(\mathfrak{g})$ skew-symmetric. Then the Lie algebras $(\mathfrak{g},\mathfrak{g}_R)$ set into duality by the inner product form a Lie bialgebra. 
\end{proposition}

Therefore, the class of dialgebras is larger than the class of bialgebras. 
Another crucial difference between the two structures is their motivation: dialgebras are related to the involutivity theorem, bialgebras related to multiplicative Poisson brackets on Lie groups\footnote{We refer to the footnote~\ref{foot:PL} for the difference between the Lie-Poisson structure and the Poisson-Lie group.}.
\begin{definition}\label{PoissonLie}
	A Poisson-Lie group is a Lie group $G$ endowed with a Poisson bracket such that the multiplication in $G$ defines a Poisson map $G \times G \to G$. 
\end{definition}
A relevant aspect of the Poisson-Lie groups is associated with their duality, that plays a key role in the generalisation of the theory of coadjoint orbits. The Poisson-Lie groups arise in correspondence of a tangent structure that takes the form of a Lie bialgebra $(\mathfrak{g},\mathfrak{g}^*)$. In particular, if $G$ is a Poisson-Lie group the pair $(\mathfrak{g},\mathfrak{g}^*)$ is a Lie bialgebra, or the tangent Lie bialgebra of $G$. The definition of Lie bialgebra is symmetric as we will see, and then if $(\mathfrak{g},\mathfrak{g}^*)$ is a Lie bialgebra, $(\mathfrak{g}^*,\mathfrak{g})$ is a Lie bialgebra as well. This will lead to the notion of a double, and then of factorisable bialgebras, that we will study in detail. We can state the following theorem:

\begin{theorem}
	Let $(\mathfrak{g},\mathfrak{g}^*)$ be a factorisable Lie bialgebra, and $G$ the Lie group $G$ associated with $\mathfrak{g}$. There is a unique multiplicative Poisson bracket on $G$ that makes it a Poisson-Lie group with tangent Lie bialgebra $(\mathfrak{g},\mathfrak{g}^*)$: the Sklyanin bracket. 
\end{theorem}
For $\varphi \in C^{\infty}(G)$ we consider the left and right gradients defined as 
\begin{equation*}
	\begin{split}
		\langle d_{\lambda}\varphi(x), X \rangle &= \left. \frac{d}{dt} \varphi(\exp(tX)\,x) \right|_{t=0} \,,\quad 
		\langle d_{\rho}\varphi(x), X \rangle = \left. \frac{d}{dt} \varphi(x\,\exp(tX)) \right|_{t=0} \,.
	\end{split}
\end{equation*}
The Sklyanin bracket for $\varphi, \psi \in C^{\infty}(G)$ is
\begin{equation}
	\{ \varphi, \psi \} = \frac{1}{2}(\langle R (d_{\rho}\varphi), d_{\rho} \psi \rangle - \langle R(d_{\lambda}\varphi), d_{\lambda} \psi \rangle  )\,,
\end{equation}
where $R$ is an $r$-matrix in the sense of $R \in \text{End} (\mathfrak{g})$.

\subsection{Lie bialgebras}
Any Lie algebra $\mathfrak{g}$ acts on the tensorial product of itself as $\mathfrak{g} \otimes \mathfrak{g} \dots \otimes \mathfrak{g}$ $p$ times as 
\begin{equation}
\begin{split} 
    X (Y_1 \otimes \dots \otimes Y_p) &= \text{ad}_X^{(p)}(Y_1 \otimes \dots \otimes Y_p) \\
    &= \text{ad}_X Y_1 \otimes \dots \otimes Y_p + \dots + Y_1 \otimes \dots \otimes \text{ad}_X Y_p\,, 
\end{split}
\end{equation}
hence for $p=2$ one has 
\begin{equation}
    \text{ad}^{(2)}_X = \text{ad}_X \otimes I + I \otimes \text{ad}_X \,, 
\end{equation}
and usually since $\text{ad}_X Y = [X,Y]$ when $Y \in \mathfrak{g}$, we write 
\begin{equation}
    (\text{ad}_X \otimes I + I \otimes \text{ad}_X)(u) = [X \otimes I + I \otimes X](u)\,, \qquad u \in \mathfrak{g} \otimes \mathfrak{g} \,. 
\end{equation}
With $\{e_i\}_{i=1}^n$ basis for $\mathfrak{g}$, any element $u \in \mathfrak{g} \otimes \mathfrak{g}$ can be written as 
\begin{equation}
    \text{ad}^{(2)}_X u = u^{ij} ([X,e_i] \otimes e_j + e_i \otimes [X,e_j]) \,, \qquad X \in \mathfrak{g}\,.
\end{equation}
The Lie algebra $\mathfrak{g}$ acts on the exterior product $\bigwedge^p \mathfrak{g}$ analogously, i.e.\ for $p=2$ and $v \in \mathfrak{g} \wedge \mathfrak{g}$
\begin{equation}
\begin{split} 
    X (Y_1 \wedge Y_2) &= [X,Y_1]\wedge Y_2 + Y_1 \wedge [X,Y_2]\,, \\ 
    X ( v^{ij} e_i \wedge e_j ) &= v^{ij} ([X,e_i]\wedge e_j+e_i \wedge [X,e_j]) \,.  
\end{split}
\end{equation}
\begin{definition}
    For non-negative integer $k$ the vector space of skew-symmetric linear mappings on $\mathfrak{g}$ with values in $M$ (vector space of a representation of $\mathfrak{g}$) is called the space of $k$-cochains on $\mathfrak{g}$ with values in $M$. 
\end{definition}
In particular, a $1$-cochain on $\mathfrak{g}$ with values in $M$ is a linear map  $\alpha \colon \mathfrak{g} \to M$, while a $0$-cochain on $\mathfrak{g}$ with values in $M$ is just an element of $M$. The coboundary of a $k$-cochain $u$ on $\mathfrak{g}$ with values in $M$ is denoted by $\delta u$, and for $k=0$ and $k=1$ they are defined as follows: 
\begin{equation}
    \begin{split}
        k&=0 \colon \quad \delta u(X) = X u \,,\quad  u \in M, ~ X \in \mathfrak{g}\,, \\
        k&=1 \colon \quad \delta v(X,Y) = X \,v(Y) - Y\, v(X) - v([X,Y]) \,, \quad v \colon \mathfrak{g}\to M, ~ X \in \mathfrak{g}\,. 
    \end{split}
\end{equation}
In particular for any $0$-cochain $u$ we have $\delta(\delta u)= 0$.

\begin{definition}
    The \emph{coboundary} of a $k$-cochain $u$ on $\mathfrak{g}$ is the $(k+1)$-cochain $\delta u$ defined by 
    \begin{equation}
    \begin{split} 
        \delta u (X_0, \dots,X_k) &= \sum_{\ell=0}^k(-1)^{\ell}X_{\ell} \, (u(X_0, \dots, \hat{X}_\ell,\dots, X_k )) \\ &~~+  \sum_{i<j}^k(-1)^{i+j} (u([X_i,X_j],X_0, \dots, \hat{X}_i,\dots, \hat{X}_j,\dots, X_k ))\,,
        \end{split}
      \end{equation}
      with $X_{\ell} \in \mathfrak{g}$ and $\hat{X}_i$ corresponds to the element that is omitted. 
\end{definition}
The property $\delta(\delta u)= 0$ can be generalised to any $k$-cochain with $k \ge 0$. 

\begin{definition}
A $k$-cochain $u$ is a \emph{$k$-cocycle}  if $\delta u=0$, and a $k$-cochain $u$ with $k \ge 0$ is a \emph{$k$-coboundary} if there exist a $(k-1)$-cochain $v$ such that $v = \delta u$. 
\end{definition}
    
Let $\mathfrak{g}$ be a Lie algebra, and $\gamma$ a linear map admitting a linear transpose $^t\gamma$ such that 
\begin{equation}
    \begin{split}
        \gamma &\colon \mathfrak{g} \to \mathfrak{g} \otimes \mathfrak{g}\,, \\
        ^t\gamma &\colon \mathfrak{g}^* \otimes \mathfrak{g}^* \to  \mathfrak{g}^*\,. 
    \end{split}
\end{equation}
Of course a linear map on $\mathfrak{g}^* \otimes \mathfrak{g}^*$ defines a bilinear map on $\mathfrak{g}^*$. 

\begin{definition}
    A Lie bialgebra $(\mathfrak{g},\mathfrak{g}^*)$ is a Lie algebra $\mathfrak{g}$ with a linear map $\gamma \colon \mathfrak{g} \to \mathfrak{g} \otimes \mathfrak{g}$ such that
    \begin{enumerate}[label=($\roman*$)]
        \item $^t \gamma \colon \mathfrak{g}^* \otimes \mathfrak{g}^* \to  \mathfrak{g}^*$ defines a Lie bracket on $\mathfrak{g}^*$ $[\,\cdot\,,\,\cdot\,]_*$ ;
        \item $\gamma$ is a $1$-cocycle on $\mathfrak{g}$ with values in $\mathfrak{g} \otimes \mathfrak{g}$, where $\mathfrak{g}$ acts on $\mathfrak{g} \otimes \mathfrak{g}$ by the adjoint representation $\text{ad}^{(2)}$. 
    \end{enumerate}
\end{definition}
The condition $(ii)$ means that the $2$-cochain $\delta\gamma$ vanishes, i.e.\
\begin{equation}\label{eq:cond_2cochain}
    \text{ad}^{(2)}_X (\gamma(Y)) - \text{ad}^{(2)}_Y (\gamma(X)) - \gamma([X,Y]) = 0\,, \qquad X,Y \in \mathfrak{g}\,. 
\end{equation}
We introduce the Lie algebra on $\mathfrak{g}^*$ as 
\begin{equation}\label{eq:Liebra_star}
    [\xi, \eta]_{*} = \,^t\gamma (\xi \otimes \eta)\,, \qquad \xi, \eta \in \mathfrak{g}^*. 
\end{equation}
Therefore, by the previous definition we can write the following 
\begin{equation}
    \langle [\xi, \eta]_{*} , X \rangle = \langle \xi \otimes \eta , \gamma(X)\rangle\,, \qquad \xi, \eta \in \mathfrak{g}^*, ~ X \in \mathfrak{g}\,. 
\end{equation}
And now we can express the~\eqref{eq:cond_2cochain} in the following alternative way,
\begin{equation}
\begin{split} 
     \langle [\xi, \eta]_{*} , [X,Y] \rangle &= \langle \xi \otimes \eta, (\text{ad}_X \otimes I + I \otimes \text{ad}_X) (\gamma(Y)) \rangle \\
     &~~-  \langle \xi \otimes \eta, (\text{ad}_Y \otimes I + I \otimes \text{ad}_Y) (\gamma(X)) \rangle\,. 
    \end{split}
\end{equation}
Considering the coadjoint action of $\mathfrak{g}$ on $\mathfrak{g}^*$ and \eqref{eq:Liebra_star}, we can also write the expression as 
\begin{equation}
\begin{split} 
     \langle [\xi, \eta]_{*} , [X,Y] \rangle &+ \langle [\text{ad}^*_X \xi, \eta]_{*} , Y \rangle + \langle [ \xi, \text{ad}^*_X \eta]_{*} , Y \rangle \\
     &-\langle [\text{ad}^*_Y \xi, \eta]_{*} , X \rangle - \langle [ \xi, \text{ad}^*_Y \eta]_{*} , X \rangle = 0\,. 
    \end{split}
\end{equation}
It is evident that the map $\gamma$ establishes a symmetry between $\mathfrak{g}$, with its bracket $[\,\cdot\,,\,\cdot\,]$ and $\mathfrak{g}^*$, with its bracket $[\,\cdot\,,\,\cdot\,]_*$. We can moreover consider the adjoint action of $\mathfrak{g}^*$ on itself and the coadjoint on $\mathfrak{g}$, so that we have 
\begin{equation}
\begin{split} 
     \langle [\xi, \eta]_{*} , [X,Y] \rangle &+ \langle \text{ad}^*_X \xi,  \text{ad}^*_{\eta}Y \rangle - \langle  \text{ad}^*_X \eta , \text{ad}^*_{\xi} Y \rangle \\
     &-\langle \text{ad}^*_Y \xi, \text{ad}^*_{\eta} X \rangle + \langle \text{ad}^*_Y \eta , \text{ad}^*_{\xi} X \rangle = 0\,. 
    \end{split}
\end{equation}
Now, we introduce the skew-symmetry map $\mu \colon \mathfrak{g}\otimes \mathfrak{g} \to \mathfrak{g}$ defining $[\,\cdot\,,\,\cdot\,]$ on $\mathfrak{g}$. Transforming the same relation again, we can see that $^t \mu \colon \mathfrak{g}^* \to \mathfrak{g}^* \otimes \mathfrak{g}^*$ is a $1$-cocycle on $\mathfrak{g}^*$ with values in $\mathfrak{g}^* \otimes \mathfrak{g}^*$, with $\mathfrak{g}^*$ acting on $\mathfrak{g}^* \otimes \mathfrak{g}^*$ by the usual adjoint action. Indeed, we have 
\begin{equation}
    \begin{split}
        \langle \,^t \mu [\xi, \eta]_{*} , X \otimes Y \rangle &- \langle  \xi,  [X, \text{ad}^*_{\eta}Y] \rangle + \langle   \eta , [X,\text{ad}^*_{\xi} Y] \rangle \\
     &+\langle  \xi, [Y,\text{ad}^*_{\eta} X ] \rangle - \langle  \eta , [Y,\text{ad}^*_{\xi} X ] \rangle = 0\,,
    \end{split}
\end{equation}
which can be rewritten as 
\begin{equation}
    \begin{split}
       \langle [\xi, \eta]_{*} , [X,Y] \rangle &=  \langle  (\text{ad}_{\xi} \otimes I + I \otimes \text{ad}_{\xi}) (^t\mu(\eta)), X \times Y \rangle \\
     &~~-  \langle  (\text{ad}_{\eta} \otimes I + I \otimes \text{ad}_{\eta}) (^t\mu(\xi)), X \times Y \rangle \,, 
    \end{split}
\end{equation}
and finally
\begin{equation}
     \text{ad}^{(2)}_{\xi} (\,^t\mu(\eta)) - \text{ad}^{(2)}_{\eta} (\,^t\mu(\xi)) - \,^t\mu([\xi,\eta]_*) = 0\,, \qquad \xi,\eta \in \mathfrak{g}^*\,. 
\end{equation} 
We can then establish the following proposition: 
\begin{proposition}
    If $(\mathfrak{g},\mathfrak{g}^*)$ with $\gamma\colon \mathfrak{g} \to \mathfrak{g}$ is a Lie bialgebra, and if $\mu$ is the Lie bracket on $\mathfrak{g}$, then $(\mathfrak{g}^*,\mathfrak{g})$ is a Lie bialgebra with $^t\mu \colon \mathfrak{g}^* \to \mathfrak{g}^* \otimes \mathfrak{g}^*$, and $^t\gamma$ is the Lie bracket on $\mathfrak{g}^*$. 
\end{proposition}

\subsection{Double of Lie bialgebras} 
Let $(\mathfrak{g},\mathfrak{g}^*)$ with $\gamma\colon \mathfrak{g}\to \mathfrak{g} \otimes \mathfrak{g}$ be a Lie bialgebra with dual $(\mathfrak{g}^*,\mathfrak{g})$ and $^t \mu \colon \mathfrak{g}^* \to \mathfrak{g}^* \otimes \mathfrak{g}^*$. Then we have the following proposition. 
\begin{proposition}
    There exists a unique Lie algebra structure on $\mathfrak{d}=\mathfrak{g} \oplus \mathfrak{g}^*$ as a vector space such that $\mathfrak{g}$ and $\mathfrak{g}^*$ are Lie subalgebras, and the natural scalar product on $\mathfrak{d}$ is invariant. 
\end{proposition}
The natural inner product $\langle\,\cdot\,,\,\cdot\,\rangle$ on $\mathfrak{g} \oplus\mathfrak{g}^*$ is 
\begin{equation}
    \langle X \,|\,Y \rangle = 0 \,, \quad  \langle \xi \,|\,\eta \rangle = 0 \,, \quad \langle X \,|\,\xi \rangle = \langle \xi, X \rangle\,, \quad X, Y \in \mathfrak{g}, ~ \xi, \eta \in \mathfrak{g}^*.
\end{equation}
We denote with $[\,\cdot\,,\,\cdot\,]_{\mathfrak{d}}$ the Lie bracket on $\mathfrak{d}$, and from the invariance of the inner product and the fact that $\mathfrak{g}$ is a Lie subalgebra, we have 
\begin{equation*}
    \langle Y | \,[ X, \xi ]_{\mathfrak{d}}\rangle = \langle [Y, X]_{\mathfrak{d}} | \, \xi \rangle = \langle [Y, X] | \, \xi \rangle  = \langle \xi, [Y, X] \rangle = \langle \text{ad}^*_X \xi \,, Y  \rangle  = \langle Y |\, \text{ad}^*_X \xi \rangle\,, 
\end{equation*}
and equivalently 
\begin{equation*}
    \langle \eta | \,[ X, \xi ]_{\mathfrak{d}}  \rangle = -\langle \eta | \, \text{ad}^*_{\xi} X \rangle  \,, 
\end{equation*}
so that we have the expression for the whole $\mathfrak{d}$ 
\begin{equation}
    [X,\xi]_{\mathfrak{d}} = \text{ad}^*_X \xi -\text{ad}^*_{\xi} X \,. 
\end{equation}
One can then prove that $[\,\cdot\,,\,\cdot\,]_{\mathfrak{d}}$ is indeed a Lie algebra. 
\begin{definition}
    The structure $\mathfrak{d}=\mathfrak{g}\oplus \mathfrak{g}^*$ built from the Lie bialgebra $(\mathfrak{g},\mathfrak{g}^*)$ is called the \emph{double} of $\mathfrak{g}$. 
\end{definition}
By construction, in the Lie algebra $\mathfrak{d}$ the subspaces $\mathfrak{g}$ and $\mathfrak{g}^*$ are complementary Lie subalgebras, and both are isotropic. 


\subsection{\boldmath \texorpdfstring{Classical Yang-Baxter equations and $r$-matrices (II)}{YB2}}
Here we will consider Lie bialgebra structures $(\mathfrak{g},\mathfrak{g}^*)$ defined by a cocycle $\delta r$, which is the coboundary of an element $r \in \mathfrak{g} \otimes \mathfrak{g}$. This is the second notion of $r$-matrices, and we will see that the CYBE for $r$ is a sufficient condition for $\delta r$ to define a Lie bracket on $\mathfrak{g}^*$. 

In the previous section we saw that a $1$-cochain in $\mathfrak{g}$ with values in $\mathfrak{g} \otimes \mathfrak{g}$ that is a coboundary of a $0$-cochain is necessarily a $1$-cocycle. Therefore, for $\gamma = \delta r$ to induce a Lie bialgebra structures, there are two more conditions: 
\begin{enumerate}
    \item $\delta r$ takes values in $\bigwedge^{\!2} \mathfrak{g}$ (i.e.\ the skew-symmetry property of $[\,\cdot\,,\,\cdot\,]_*$); 
    \item the Jacobi identity for $[\,\cdot\,,\,\cdot\,]_*$ defined for $\delta r$ must be satisfied. 
\end{enumerate}
We introduce $a$, $s$ to identify respectively the skew-symmetric and symmetric part of $r$: 
\begin{equation}
    r = a + s\,, \qquad a \in {\bigwedge}^{\!2} \mathfrak{g}\,,~ s \in S^2 \mathfrak{g}\,. 
\end{equation}
We can always look at an element $r \in \mathfrak{g} \otimes \mathfrak{g}$ as a bilinear form on $\mathfrak{g}^*$, and so for any element in $\mathfrak{g} \otimes \mathfrak{g}$ we associate the map $\bar{r} \colon \mathfrak{g}^* \to \mathfrak{g} $ such that (in two notations)
\begin{equation}\label{eq:r_bar_map}
    \bar{r}(\xi)(\eta) = r(\xi , \eta) = \langle \eta, \bar{r}\xi \rangle\,,\qquad \xi, \eta \in \mathfrak{g}^*\,,
\end{equation}
so that an element $\bar{r}(\xi)$ is seen as a linear form on $\mathfrak{g}^*$. 
We introduce the transpose of the map as $^t \bar{r} \colon \mathfrak{g}^* \to \mathfrak{g}$. Then we have 
\begin{equation}
    \bar{a} = \frac{1}{2}(\bar{r}-\,^t \bar{r})\,, \qquad \bar{s} = \frac{1}{2}(\bar{r}+\, ^t\bar{r})\,. 
\end{equation}
We introduce $\gamma = \delta r$, and by definition we have 
\begin{equation}
    \gamma(X) = \text{ad}^{(2)}_X r = (\text{ad}_X \otimes I + I \otimes \text{ad}_X)(r)\,, \qquad X \in \mathfrak{g}\,m 
\end{equation}
where $r=r^{ij}e_i \otimes e_j$ for $\{e_i\}_{i=1}^n$ basis for $\mathfrak{g}$. The previous expression can be also written as 
\begin{equation}
	\begin{split}
		\text{ad}^{(2)}_X r &= r^{ij}(\text{ad}_X e_i \otimes e_j + e_i \otimes \text{ad}_X e_j) \\
		&= [X \otimes I + I \otimes X, r] \,. 
	\end{split}
\end{equation}
In the precious section we introduced $[\xi,\eta]_*\!=\!\,^t \gamma(\xi,\eta)$ for $\xi, \eta \in \mathfrak{g}^*$, and when $\gamma = \delta r$ we introduce the notation $[\,\cdot\,,\,\cdot\,]_r$ for the bracket on $\mathfrak{g}^*$. 

We have that the condition $(i)$ of skew-symmetry for the bracket is true if and only if the symmetric part of $r$ (i.e.\ $s$) is invariant under the adjoint action 
\begin{equation}
	\text{ad}^{(2)}_X s = 0\,, \qquad X \in \mathfrak{g}\,,  
\end{equation}
also written as $[X \otimes I + I \otimes X, s]= 0$. We have that whenever $s$ is $\text{ad}$-invariant, then $\delta r = \delta a$. This condition is satisfied trivially when $s=0$, i.e.\ when $r$ is skew-symmetric ($r=a$). 

\begin{proposition}
	If $r$ is skew-symmetric then the Lie bracket $[\,\cdot\,,\,\cdot\,]_r$ on $\mathfrak{g}^*$ has the expression 
	\begin{equation}\label{eq:r_bracket}
		[\xi, \eta]_r = \text{ad}^*_{\bar{r}\xi} \eta - \text{ad}^*_{\bar{r}\eta} \xi\,, \qquad \xi, \eta \in \mathfrak{g}^*\,. 
	\end{equation}
\end{proposition} 
Indeed, from the definition of $\delta r$ we have with $X \in \mathfrak{g}$
\begin{equation}
	\begin{split} 
	\langle \,^t (\delta r) (\xi, \eta),X \rangle &= ((\text{ad}_X \otimes I + I \otimes \text{ad}_X)(r))(\xi,\eta) \\ 
	&= -r(\text{ad}^*_{X}\xi,\eta) - r(\xi, \text{ad}^*_X \eta)   \,.
	\end{split} 
\end{equation}
With the skew-symmetry of $r$ and the definition of $\bar{r}$ we get 
\begin{equation}
	\begin{split}
		-r(\text{ad}^*_{X}\xi,\eta) - r(\xi, \text{ad}^*_X \eta)  &= r(\eta,\text{ad}^*_{X}\xi) - r(\xi, \text{ad}^*_X \eta)  \\
		&= \bar{r}(\eta)(\text{ad}^*_{X}\xi) - \bar{r}(\xi)( \text{ad}^*_X \eta) \\
		&= \langle \text{ad}^*_X \xi, \bar{r}\eta \rangle - \langle \text{ad}^*_X \eta, \bar{r}\xi \rangle \,.
	\end{split}
\end{equation}
From the properties of the coadjoint and adjoint action we have 
\begin{equation}
\begin{split}
	\langle \text{ad}^*_X \xi, \bar{r}\eta \rangle - \langle \text{ad}^*_X \eta, \bar{r}\xi \rangle &= - \langle  \xi, \text{ad}_X \bar{r}\eta \rangle + \langle \eta, \text{ad}_X \bar{r}\xi \rangle \\
	&=  \langle  \xi, \text{ad}_{\bar{r}\eta} X  \rangle - \langle \eta, \text{ad}_{\bar{r}\xi} X  \rangle \\
	&= -\langle  \text{ad}^*_{\bar{r}\eta} \xi,  X  \rangle + \langle \text{ad}^*_{\bar{r}\xi} \eta,  X  \rangle \,,
\end{split}
\end{equation}
and for the arbitrariness of $X \in \mathfrak{g}$ we obtain \eqref{eq:r_bracket}. 

To show the Jacobi identity in condition $(ii)$ we make use of the Schouten bracket, defined for an element $r \in \bigwedge^2 \mathfrak{g}$ with itself and denoted with $\SB{r,r}$. This is the element in $\bigwedge^3 \mathfrak{g}$ defined by 
\begin{equation}\label{eq:schouten_r}
	\SB{r,r}(\xi,\eta,\zeta) = -2( \langle \zeta, [\bar{r}\xi,\bar{r}\eta] \rangle + \langle \xi, [\bar{r}\eta,\bar{r}\zeta] \rangle + \langle \eta, [\bar{r}\zeta,\bar{r}\xi] \rangle ) \,, 
\end{equation}
\begin{proposition}
	A necessary condition for $\gamma=\delta r$ with $r \in \bigwedge^2 \mathfrak{g}$ to define a Lie bracket on $\mathfrak{g}^*$ is that $\SB{r,r} \in \bigwedge^3 \mathfrak{g}$ is ad-invariant. 
\end{proposition}
The element $\SB{r,r}$ is a $0$-cochain on $\mathfrak{g}$ with values in $\bigwedge^3 \mathfrak{g}$, and it is ad-invariant if and only if $\delta(\SB{r,r})=0$. Hence, the proposition will follow from 
\begin{equation}\label{eq:Jacobi_for_r}
	\begin{split} 
	-\frac{1}{2}\delta(\SB{r,r})(X)(\xi,\eta,\zeta) &= (\langle [[\xi,\eta]_r,\zeta]_r,X \rangle + \langle [[\eta,\zeta]_r,\xi]_r,X \rangle\\ 
	&~~+ \langle [[\zeta,\xi]_r,\eta]_r,X \rangle) \,,
\end{split} 
\end{equation}
for $\xi, \eta, \zeta \in \mathfrak{g}^*$, $X \in \mathfrak{g}$, and with $1/2$ a conventional coefficient.  For each of the three terms in the right hand side of~\eqref{eq:Jacobi_for_r} we have the following 
\begin{equation}\label{eq:Jabobi_for_r_step}
	\begin{split}
		\langle [[\xi,\eta]_r,\zeta]_r,X \rangle &= \langle [\text{ad}^*_{\bar{r} \xi} \eta - \text{ad}^*_{\bar{r} \eta} \xi,\zeta]_r,X \rangle  \\
		&= \langle \text{ad}^*_{\bar{r}(\text{ad}^*_{\bar{r} \xi} \eta - \text{ad}^*_{\bar{r} \eta} \xi)} \zeta,X \rangle - \langle \text{ad}^*_{\bar{r}\zeta} (\text{ad}^*_{\bar{r} \xi} \eta - \text{ad}^*_{\bar{r} \eta} \xi)  ,X \rangle  \\ 
		&= - \langle \text{ad}^*_{\bar{r}(\text{ad}^*_{\bar{r} \xi} \eta - \text{ad}^*_{\bar{r} \eta} \xi)} \zeta,X \rangle\,.
	\end{split}
\end{equation} 
Using the skew-symmetry of $r$ and the relation 
\begin{equation}
	\text{ad}^*_X\,\text{ad}^*_Y - \text{ad}^*_Y\,\text{ad}^*_X = \text{ad}^*_{[X,Y]}\,, 
\end{equation}
for $Y=\bar{r}\xi$ in \eqref{eq:Jabobi_for_r_step}, we get 
\begin{equation*}
	\begin{split}
		\langle [[\xi,\eta]_r,\zeta]_r,X \rangle &= \langle \bar{r}\,\text{ad}^*_{X} \zeta, \text{ad}^*_{\bar{r}\xi} \eta \rangle - \langle \bar{r}\,\text{ad}^*_{X} \zeta, \text{ad}^*_{\bar{r}\eta} \xi \rangle \\
		&~~+ \langle \text{ad}^*_{\bar{r}\xi}\,\text{ad}^*_X \eta , \bar{r}\zeta \rangle + \text{ad}^*_{\bar{r}\xi}\,\text{ad}^*_{[X,\bar{r}\xi]} \eta , \bar{r}\zeta \rangle - \text{ad}^*_{X}\,\text{ad}^*_{\bar{r}\eta} \xi , \bar{r}\zeta \rangle \,,
	\end{split}
\end{equation*}
and finally we have 
\begin{equation}\label{eq:Jacobi_for_r_last}
	\begin{split}
		\langle [[\xi,\eta]_r,\zeta]_r,X \rangle &= \langle \eta, [\bar{r}\,\text{ad}^*_X\zeta, \bar{r}\xi] \rangle + \langle \xi, [\bar{r}\eta,\bar{r}\,\text{ad}^*_X\zeta] \rangle + \langle \text{ad}^*_X \eta, [\bar{r}\zeta, \bar{r}\xi] \rangle \\
		&~~+\langle \text{ad}^*_X\,\text{ad}^*_{\bar{r}\zeta}\eta , \bar{r}\xi \rangle - \langle \text{ad}^*_X\,\text{ad}^*_{\bar{r}\eta}\xi , \bar{r}\zeta \rangle\,. 
	\end{split}
\end{equation}
When considered in the cyclic sum of the variables $\xi, \eta, \zeta$, the terms appearing in the second row of~\eqref{eq:Jacobi_for_r_last} vanish. Finally, we consider the left hand side of~\eqref{eq:Jacobi_for_r}, and we have 
\begin{equation}
	\begin{split}
	\delta(\SB{r,r})(X)(\xi,\eta,\zeta) &= \text{ad}^{(3)}_X\SB{r,r}(\xi,\eta,\zeta) \\
	&= -\SB{r,r}(\xi,\eta,\text{ad}^*_X \zeta) -\SB{r,r}(\eta,\zeta,\text{ad}^*_X \xi) -\SB{r,r}(\zeta,\xi,\text{ad}^*_X \eta) \,,
	\end{split}
\end{equation}
and for one of the three elements at the right hand side we have from~\eqref{eq:schouten_r}
\begin{equation}
	-\SB{r,r}(\xi,\eta,\text{ad}^*_X \zeta) = 2( \langle \text{ad}^*_X\zeta, [\bar{r}\xi,\bar{r}\eta] \rangle + \langle \xi, [\bar{r}\eta,\bar{r}\text{ad}^*_X \zeta] \rangle + \langle \eta, [\bar{r}\text{ad}^*_X \zeta,\bar{r}\xi] \rangle )\,.
\end{equation} 
The two expressions for the cyclic sum on $(\xi,\eta,\zeta)$ coincide, hence the bilinear form induced by $\gamma = \delta r$ on $\mathfrak{g}^*$ is indeed a Lie bracket. 

If $r \in \mathfrak{g} \otimes \mathfrak{g}$ is skew-symmetric, then the condition determining the ad-invariance for $\SB{r,r}$ (i.e.\ $\delta (\SB{r,r}) = 0$) takes the name of generalised CYBE. A sufficient condition for the ad-invariance of $\SB{r,r}$ is 
\begin{equation}\label{eq:YB_second_red}
	\SB{r,r}= 0\,,
\end{equation}
and this corresponds to a specific case of CYBE, as we will see.  

\begin{definition}
	Let $r \in \mathfrak{g} \otimes \mathfrak{g}$ given in terms of its skew-symmetric and symmetric part  $r=a+s$. If $s$ and $\SB{a,a}$ are ad-invariant $r$ is an $r$-matrix. If moreover $r$ is skew-symmetric $r=a$ and $\SB{a,a}=0$ then $r$ is called a \emph{triangular} $r$-matrix. 
\end{definition}

Usually in literature is more common to encounter these expressions in the tensor formalism, that we will introduce in the next section. In particular, as we will see, in the tensor formalism the symmetric part of the $r$ matrix is directly introduced as ad-invariant. 

It is worth noticing that other authors define the $r$-matrix in a slightly different way (see e.g.\ \cite{BabVia35KS,LiParm43KS,Reim26KS}), where the symmetric part of $r$ (i.e.\ $s$) is not necessarily ad-invariant, and then the corresponding Lie bracket is not in general inducing a Lie bialgebra $(\mathfrak{g}^*,\mathfrak{g})$ on the dual. 

Let $r \in \mathfrak{g} \otimes \mathfrak{g}$, we introduce $\overline{\langle r,r\rangle} \colon \bigwedge^2 \mathfrak{g}^* \to \mathfrak{g}$ defined as 
\begin{equation}
	\overline{\langle r,r\rangle} (\xi, \eta) = [\bar{r}\xi, \bar{r}\eta] - \bar{r}[\xi, \eta]_r\,, 
\end{equation} 
and the $\langle r , r \rangle$ defined accordingly as
\begin{equation}
	\RR{r,r}(\xi, \eta, \zeta) = \langle \zeta, \overline{\langle r,r\rangle} (\xi, \eta)  \rangle\,,
\end{equation}
hence there is an identification of the map $\overline{\langle r,r\rangle}$ with an element of $\bigwedge^2 \mathfrak{g} \otimes \mathfrak{g}$. In particular, whenever the symmetric part of $r$ is ad-invariant $\RR{r,r} \in \bigwedge^3 \mathfrak{g}$. Indeed, we have the following theorem. 

\begin{theorem}Let $r = a+s$ an element of $\mathfrak{g} \otimes \mathfrak{g}$, where $a$ is skew-symmetric and $s$ symmetric and ad-invariant. We have the following: 
	\begin{enumerate}[label=\textup{(}\roman*\textup{)}]
			\item $\RR{a,a}$ is in $\bigwedge^3 \mathfrak{g}$ and it is given by 
		\begin{equation}
			\RR{a,a} = - \frac{1}{2}\SB{a,a}\,;
		\end{equation}
	\item $\RR{s,s}$ is an ad-invariant element in $\bigwedge^3\mathfrak{g}$ and 
	\begin{equation} 
		\overline{\langle s,s\rangle} = [\bar{s}\xi, \bar{s}\eta]
	\end{equation}
		\item $\RR{r,r}$ is in $\bigwedge^3 \mathfrak{g}$ and 
		\begin{equation}
			\RR{r,r} = \RR{a,a} + \RR{s,s}
		\end{equation}
	\end{enumerate}

\end{theorem} 
If $r \in \mathfrak{g} \otimes \mathfrak{g}$ with $r=a+s$, where $a$ is skew-symmetric and $s$ symmetric and ad-invariant, a sufficient condition for $\SB{a,a}$ to be ad-invariant is 
\begin{equation}\label{eq:YB_second}
	\RR{r,r} = 0\,, 
\end{equation}
and one says that a $r \in \mathfrak{g} \otimes \mathfrak{g}$ with symmetric part ad-invariant and satisfying~\eqref{eq:YB_second} is a classical $r$-matrix. The equation \eqref{eq:YB_second} is again a CYBE and when written in the form 
\begin{equation}
	\RR{a,a} = - \RR{s,s}\,, 
\end{equation}
is the mCYBE. 

\begin{definition}
	A $r$-matrix satisfying \eqref{eq:YB_second} is called \emph{quasi-triangular}. If moreover the symmetric part of $r$ is invertible, $r$ is called \emph{factorisable}. 
\end{definition}
When $r$ is skew-symmetric, the \eqref{eq:YB_second} reduces to \eqref{eq:YB_second_red}. Triangular, quasi-triangular and factorisable $r$-matrices give rise to coboundary Lie bialgebras. Moreover, we can now establish a proper connection with the formalism of the $R$-matrix treated in the previous section. 

Since $r=a+s$, then its inverse is $^t r = -a +s $, and if $r$ is a solution to CYBE, then $^t r$ is a solution as well. We introduce 
\begin{equation}
    r_+ = r\,, \qquad r_- = - ^t r \,,
\end{equation}
and an $r$-matrix is quasi-triangular if and only if the maps the maps $\bar{r}_+$, $\bar{r}_-$ as in \eqref{eq:r_bar_map} 
\begin{equation}
    \bar{r}_+ = \bar{a} + \bar{s} \,, \qquad \bar{r}_- = \bar{a} - \bar{s}
\end{equation}
are Lie algebra morphism from $(\mathfrak{g}^*, \mathfrak{g})$ with  $[\,\cdot\,,\,\cdot\,]_r$ \eqref{eq:r_bracket} to $\mathfrak{g}$. 

\begin{proposition}
    Let us consider $r=a+s$ an element of $\mathfrak{g} \otimes \mathfrak{g}$ and a factorisable $r$-matrix. Then, there exists $R \colon \mathfrak{g} \to \mathfrak{g}$ such that the composition $R \circ \bar{s} \colon \mathfrak{g}^* \to \mathfrak{g}$ is skew-symmetric and in particular 
    \begin{equation}
        \bar{a} = R \circ \bar{s} \,.
    \end{equation}
    Then, we have 
    \begin{equation}
        R = \bar{a} \circ \bar{s}^{-1}\,, \qquad r = (R+I) \circ \bar{s} 
    \end{equation}
    and in brackets 
    \begin{equation}
        [R(X), Y] + [X,R(Y)] = \bar{a}[ \bar{s}^{-1}X\,, \bar{s}^{-1}Y ]_{a}\,. 
    \end{equation}
\end{proposition}

\subsection{Tensor formalism}
As already noticed, a way to express $\delta r(X)$ is via the tensor formalism, i.e.\ with $r \in \mathfrak{g} \otimes \mathfrak{g}$, $X \in \mathfrak{g}$ we have 
\begin{equation}
	\delta r(X) = [\,X \otimes I + I \otimes X, r]\,. 
\end{equation}
For $r \in \mathfrak{g} \otimes \mathfrak{g}$ and we introduce the notation $r_{12}$, $r_{13}$, $r_{23}$ for the elements in $\mathfrak{g} \otimes \mathfrak{g} \otimes \mathfrak{g}$ as: 
\begin{equation}
	r_{12}=r \otimes I\,, \qquad r_{23}= I \otimes r\,, 
\end{equation}
and the third element is 
\begin{equation}
	r_{13} = \sum_k u_k \otimes I \otimes  v_k \,,  \qquad r=\sum_k u_k \otimes  v_k\,. 
\end{equation}
Then we construct the elements in $\mathfrak{g} \otimes \mathfrak{g} \otimes \mathfrak{g}$ as 
\begin{align*}
    [r_{12},r_{13}] &= [ \sum_{k}  u_k  \otimes  v_k \otimes I, \sum_{\ell}  u_\ell \otimes I \otimes  v_\ell ] = \sum_{k,\ell} [u_k,u_\ell] \otimes v_k \otimes  v_\ell \,,\\
     [r_{12},r_{23}] &= [ \sum_{k}  u_k  \otimes  v_k \otimes I, \sum_{\ell}  u_\ell \otimes I \otimes  v_\ell ] = \sum_{k,\ell}  u_k \otimes [v_k,u_\ell] \otimes  v_\ell \,, \\
         [r_{13},r_{23}] &= [ \sum_{k}  u_k  \otimes  v_k \otimes I, \sum_{\ell}  u_\ell \otimes I \otimes  v_\ell ] = \sum_{k,\ell} u_k \otimes u_\ell \otimes [v_k, v_\ell]\,. \\   
\end{align*}
In this notation the symmetric part of $r$ is directly ad-invariant, then 
\begin{align}
	\RR{r,r}&= [r_{12},r_{13}] +  [r_{12},r_{23}] +  [r_{13},r_{13}]\,, \\
	\RR{s,s} &= [s_{12},s_{13}] =  [s_{12},s_{23}] =  [s_{13},s_{13}]\,.
\end{align}
Indeed, we can construct the previous elements in the map formalism and check the result. We have 
\begin{align*}
	[r_{12},r_{13}](\xi, \eta, \zeta) &= \langle \xi , [\,^t \bar{r}\eta , \,^t \bar{r} \zeta  ] \rangle \,,\\  
	[r_{12},r_{23}](\xi, \eta, \zeta) &= \langle \eta , [\,^t \bar{r}\zeta , \,^t \bar{r} \xi  ] \rangle \,,  \\  
	[r_{13},r_{23}](\xi, \eta, \zeta) &= \langle \zeta , [\,^t \bar{r}\xi , \,^t \bar{r} \eta  ] \rangle \,.
\end{align*}
For the definition of $\RR{r,r}$ we have 
\begin{equation}
	\begin{split}
		\RR{r,r} (\xi,\eta,\zeta) &= \langle \zeta, [\bar{r}\xi, \bar{\eta}] \rangle - \langle \zeta, \bar{r}(\text{ad}^*_{\bar{r}\xi} \eta + \text{ad}^*_{\bar{r}\eta} \xi )  \rangle \,, 
	\end{split}
\end{equation}
and the ad-invariance of $s$ in $r=s+a$ gives 
\begin{equation}
	\text{ad}^*_{\bar{r}\xi} \eta + \text{ad}^*_{\bar{r}{\eta}} \xi = \text{ad}^*_{\bar{a}\xi} \eta + \text{ad}^*_{\bar{a}{\eta}} \xi = [\xi, \eta]_r\,. 
\end{equation}
With this we verify the equality and we can establish the form of CYBE in tensor notation, i.e.\
\begin{equation}
	[r_{12},r_{13}] +  [r_{12},r_{23}] +  [r_{13},r_{13}]=0\,. 
\end{equation}

Let $(\mathfrak{g},\mathfrak{g}^*)$ be a Lie bialgebra structure defined by an $r$-matrix $r \in \mathfrak{g} \otimes \mathfrak{g}$. In this case $\mathfrak{g}_*$ has a Lie structure $[\,\cdot\,,\,\cdot\,]_r$ given by 
\begin{equation}
    \langle [\xi, \eta]_r , X \rangle = [X \otimes I + I \otimes X, r](\xi, \eta) \,, \qquad X \in \mathfrak{g}\,.  
\end{equation}
Since $\mathfrak{g}$ as a vector space can be identified as the dual of $\mathfrak{g}^*$ (i.e.\ $(\mathfrak{g}^*)^* \simeq \mathfrak{g}$), then $\mathfrak{g}$ has the linear Poisson structure $\{\,\cdot\,,\,\cdot\,\}_r$ defined as 
\begin{equation}
    \{\,\cdot\,,\,\cdot\,\}_r(X) = \langle X , [\xi, \eta]_r \rangle\,,
\end{equation}
from which we have 
\begin{equation}
    \{\xi, \eta\}_r(X) = [X \otimes I + I \otimes X, r ] (\xi,\eta)\,. 
\end{equation}
In matrix representation $L \in \mathfrak{g}=\mathfrak{gl}(n)$. Then, with $I$ identity matrix of order $n$, the elements $L \otimes I$ and $I \otimes L$ are matrices of order $n^2$. With $L= (a_{ij})_{i,j=1}^n$ we have the following representation of the two elements 
\begin{align*}
 L \otimes I&= \begin{pmatrix}
a_{11} & 0 & \cdots & 0 & \cdots & a_{1n} & 0 & \cdots & 0 \\
0 & a_{11} & \cdots & 0 & \cdots & 0 & a_{1n} & \cdots & 0 \\
\vdots & \vdots & \ddots & \vdots & \ddots & \vdots & \vdots & \ddots & \vdots \\
0 & 0 & \cdots & a_{11} & \cdots & 0 & 0 & \cdots & a_{1n} \\
\vdots & \vdots & \vdots & \vdots & \vdots & \vdots & \vdots & \vdots & \vdots \\
a_{n1} & 0 & \cdots & 0 & \cdots & a_{nn} & 0 & \cdots & 0 \\
0 & a_{n1} & \cdots & 0 & \cdots & 0 & a_{nn} & \cdots & 0 \\
\vdots & \vdots & \ddots & \vdots & \ddots & \vdots & \vdots & \ddots & \vdots \\
0 & 0 & \cdots & a_{n1} & \cdots & 0 & 0 & \cdots & a_{nn}
\end{pmatrix} \\[1ex]
  I \otimes L &= 
\begin{pmatrix}
L & 0 & \cdots & 0 \\
0 & L & \cdots & 0 \\
\vdots & \vdots & \ddots & \vdots \\
0 & 0 & \cdots & L
\end{pmatrix}
\end{align*}
The Poisson structure of $\mathfrak{g}$ is then completely specified by knowing the Poisson brackets of the coordinate functions on $\mathfrak{g}$. It is then enough to know the Poisson bracket for the coefficients $\{a_{ij},a_{k\ell}\}$ where $a_{ij}$ for fixed indices $i,j$ is considered as the linear function on $\mathfrak{g}$ which associate its coefficients with the $i$-th row and $j$-th column of the matrix $L \in \mathfrak{g}$. Given that, these brackets can be considered as the elements of a matrix of order $n^2$, usually denoted by $\{L \overset{\otimes}{,} L\}$ which then has the form 
\begin{equation*}
 \{L \overset{\otimes}{,} L\} =    \begin{pmatrix}
\{a_{11}, a_{11}\} & \cdots & \{a_{11}, a_{1n}\} & \{a_{12}, a_{11}\} & \cdots & \{a_{1n}, a_{1n}\} \\
\{a_{11}, a_{21}\} & \cdots & \vdots & \vdots & \ddots & \vdots \\
\vdots & \ddots & \vdots & \vdots & \ddots & \vdots \\
\{a_{11}, a_{n1}\} & \cdots & \{a_{11}, a_{nn}\} & \vdots & \ddots & \vdots \\
\{a_{21}, a_{11}\} & \cdots & \{a_{21}, a_{1n}\} & \vdots & \ddots & \vdots \\
\vdots & \ddots & \vdots & \vdots & \ddots & \vdots \\
\{a_{21}, a_{n1}\} & \cdots & \{a_{21}, a_{nn}\} & \vdots & \ddots & \vdots \\
\vdots & \ddots & \vdots & \vdots & \ddots & \vdots \\
\{a_{n1}, a_{n1}\} & \cdots & \{a_{n1}, a_{nn}\} & \cdots & \cdots & \{a_{nn}, a_{nn}\}
\end{pmatrix}
\end{equation*}
Then, considering the element $[L \otimes I + I\otimes L,r] \in \mathfrak{g} \otimes \mathfrak{g}$ on the pair $(ij, k\ell)$ corresponds to taking the $(ij, k\ell)$ element of $\{L \overset{\otimes}{,} L\} $, and then we can write 
\begin{equation}
    \{L \overset{\otimes}{,} L\} = [L \otimes I + I\otimes L,r]\,, 
\end{equation}
which is also known as the first Russian formula. 

\begin{exercise}
	Let $\mathfrak{g}$ be the Lie algebra of dimension 2 with basis $X, Y$ and commutation relation
	\begin{equation}
		[X,Y] = X\,.
	\end{equation}
Show that $r=X \wedge Y$ is a triangular $r$-matrix. 
\end{exercise}

\begin{exercise}
	On $\mathfrak{g}=\mathfrak{sl}(2,\mathbb{C})$ with the basis given by $H,X,Y$ where 
	\begin{equation}
		H= \begin{pmatrix}
			1 & 0 \\
			0 & -1
		\end{pmatrix} \qquad X = \begin{pmatrix}
		0 & 1 \\
		0 & 0
	\end{pmatrix} \qquad Y = \begin{pmatrix}
	0 & 0 \\
	1 & 0
\end{pmatrix}
	\end{equation}
and commutation relations 
\begin{equation}\label{eq:sl2basis}
	[H,X] = 2X\,, \qquad [H,Y] = -2Y\,, \qquad [X,Y]=H\,. 
\end{equation} 
On $\mathfrak{g}^*$ there is the Lie bracket $[\,\cdot\,,\,\cdot\,]_*$ with commutation relations 
\begin{equation}\label{eq:sl2cstar}
	[H^*,X^*]_* = \frac{1}{4}X^*\,, \qquad [H^*,Y^*]_* = \frac{1}{4}X^* \, \qquad [X^*,Y^*]_* = 0\,. 
\end{equation}
Show that the $r$-matrix defined as 
\begin{equation}
	r= \frac{1}{8}(H \otimes H + 4 X \otimes Y)
\end{equation}
is factorisable, and that $[\,\cdot\,,\,\cdot\,]_r$ reproduces $[\,\cdot\,,\,\cdot\,]_*$ in~\eqref{eq:sl2cstar}. 
\end{exercise}

\begin{exercise}
	On $\mathfrak{g}=\mathfrak{sl}(2,\mathbb{C})$ show that the $r$-matrix defined as 
	\begin{equation}
		r = X \otimes H - H \otimes X 
	\end{equation}
	is triangular and evaluate $\delta r$ on the basis $X,Y,H$ as in~\eqref{eq:sl2basis}. 
\end{exercise}

\section{Toda chains}\label{sec:toda}
We consider in this part different ways of reproducing the celebrated Toda chains by involving the formalism here developed. 

\subsection{\boldmath Open Toda chain with a non-skew symmetric $r$-matrix~\cite{CauDelSin}}
We recall that the natural framework to define our phase space is a coadjoint orbit of $G_R$ in $\mathfrak{g}^*$
\begin{equation} \label{eq:coadj_orbit}
    {\mathcal O}_\Lambda=\{\text{Ad}^{R*}_{g}\cdot \Lambda;g\in G_R\}\,,~~\text{for some}~~\Lambda\in\mathfrak{g}^*\,. 
\end{equation}
Let us consider $\mathfrak{g}=\mathfrak{sl}(N+1)$, the Lie algebra of $(N+1)\times (N+1)$ traceless real matrices, $\mathfrak{g}_+$ the Lie subalgebra of skew-symmetric matrices and $\mathfrak{g}_-$ the Lie subalgebra of upper triangular traceless matrices, yielding
	\begin{equation}
	\mathfrak{g}=\mathfrak{g}_+\oplus \mathfrak{g}_-\,.    
	\end{equation}
	Here $R=P_+-P_-$ and $R_\pm=\pm P_\pm$ with $P_\pm$ the projector on $\mathfrak{g}_\pm$ along $\mathfrak{g}_\mp$. 
	We introduce the inner product as 
	\begin{equation} 
 \label{bilinear_form}
	\langle\,X|Y\rangle={\rm Tr}(XY)
	\end{equation} 
	allows the identification $\mathfrak{g}^*\simeq \mathfrak{g}$, and it induces the decomposition  
	\begin{equation}
	\mathfrak{g}^*=\mathfrak{g}_-^*\oplus \mathfrak{g}_+^*\simeq\mathfrak{g}_+^\perp\oplus \mathfrak{g}_-^\perp\,,
	\end{equation}
	where $\mathfrak{g}_\pm^\perp$ is the orthogonal complement of $\mathfrak{g}_{\pm}$ with respect to $\langle\,~,~\rangle$: $\mathfrak{g}_+^\perp$ is the subspace of traceless symmetric matrices and $\mathfrak{g}_-^\perp$ the subspace of strictly upper triangular matrices. Let us choose as $\Lambda$ in \eqref{eq:Adjoint_L}  the following 
	\begin{equation}  \label{eq:Lambda}
	\Lambda =\begin{pmatrix}
		0 & 1 & 0 & 0 &\dots & 0\\
		1 & 0 & 1 & 0 &\dots & 0\\
		0 & 1 & 0 & 1 &\dots & 0\\
		0 & 0 & 1 &\ddots &\ddots & \vdots\\
		\vdots & & & \ddots & \ddots & 1\\
		0 & 0 & 0 & \dots &1 & 0\\
	\end{pmatrix}\in \mathfrak{g}_-^*\simeq\mathfrak{g}_+^\perp
	\end{equation} 
	and consider its orbit under the (co)adjoint action of $G_-$, the Lie subgroup associated to $\mathfrak{g}_-$ consisting of upper triangular matrices with unit determinant. 
	
This corresponds to the particular case where $L$ is defined with respect to $g_- \in G_-$ so that 
	\begin{equation*}
		L=\text{Ad}^{R*}_{g} \cdot \Lambda=-R_-^*(\text{Ad}^*_{g_-}\cdot \Lambda),
	\end{equation*}
	and the coadjoint orbit ${\mathcal O}_{\Lambda}$ lies in $\mathfrak{g}_-^*$. 
	Using $\langle\,\cdot~|~\cdot\,\rangle$ we can identify the adjoint and coadjoint actions. Also, we use it to identify the transpose $A^*:\mathfrak{g}^*\to\mathfrak{g}^*$ of any linear map $A:\mathfrak{g}\to\mathfrak{g}$ with the transpose of $A$ with respect to $\langle\,\cdot~|~\cdot\,\rangle$ defined on $\mathfrak{g}$. Writing $\langle \xi,X \rangle=\langle Y|X\rangle$, this means that we have
	$$\langle A^*(\xi),X\rangle =\langle\xi,A(X)\rangle =\langle Y|A(X) \rangle=\langle A^*(Y)|X \rangle\,.$$
	This allows us to work with 
	\begin{equation}
		L=-R_-^*(g_-\, \Lambda\,g_-^{-1})=-R_-^*(g\, \Lambda\,g^{-1}) \,,
	\end{equation}
	where we have dropped the redundant subscript on $g$ in the second equality with $g=g_-\in G_-$.
From the definitions $\langle\,X\,|\,R_\pm Y\,\rangle=\langle\,R^*_\pm X\,|\,Y\,\rangle$ and  $\langle\,X\,|\,P_\pm Y\,\rangle=\langle\,\Pi_\mp X\,|\,Y\,\rangle$, where we denote by $\Pi_\pm$ the projector onto $\mathfrak{g}_\pm^\perp$ along $\mathfrak{g}_\mp^\perp$, we find $R^*_\pm=\pm\Pi_\mp\,$. Note that this is an example of non-skew-symmetric $r$-matrix since
 \begin{equation}
     R^*=\Pi_--\Pi_+\neq -R=P_--P_+\,.
 \end{equation}
	Now, $g\,\Lambda \,g^{-1}$ is of the form
	\begin{equation}
	g\,\Lambda \,g^{-1}=\begin{pmatrix}
		~a_1~ & * & * & * &\dots & *\\[1ex]
		b_1 & a_2 & * & * &\dots & *\\[1ex]
		0 & b_2 & a_3 & * &\dots & *\\[1ex]
		0 & 0 & b_3 &\ddots &\ddots & \vdots\\[1ex]
		\vdots & & & \ddots & \ddots & *\\[1ex]
		0 & 0 & 0 & \dots &b_{N} & ~a_{N+1}~\\
	\end{pmatrix}\,.
	\end{equation} 
So, we find
	\begin{equation}
	\label{form_L_Flaschka}
	L=\Pi_+(g\,\Lambda \,g^{-1})=\begin{pmatrix}
		~a_1~ & b_1 & 0 & 0 &\dots & 0\\[1ex]
		b_1 & a_2 & b_2 & 0 &\dots & 0\\[1ex]
		0 & b_2 & a_3 & b_3 &\dots & 0\\[1ex]
		0 & 0 & b_3 &\ddots &\ddots & \vdots\\[1ex]
		\vdots & & & \ddots & \ddots & b_{N}\\[1ex]
		0 & 0 & 0 & \dots &b_{N} & ~a_{N+1}~\\
	\end{pmatrix}\,,
	\end{equation} 
	i.e.\ it is symmetric tridiagonal. The Casimir functions $H_{\ell}$ and their differential $\nabla$ with respect to the inner product $\langle\,\cdot~|~\cdot\, \rangle$ are 
    \begin{equation}
        H_{\ell}(L) = \frac{1}{\ell+1} \text{tr} L^{\ell+1}\,, \qquad \nabla H_{\ell} = \text{tr}L^{\ell}\,. 
    \end{equation}
    We select the first one, i.e.
	\begin{equation} \label{eq:Ham_1_Toda}
		H_1(L) = \frac{1}{2} \,\text{tr}\, L^2\,, 
	\end{equation}
	and the corresponding Lax equation is 
	\begin{equation} 
	R_+\nabla H_1(L)=P_+(L)=\begin{pmatrix}
		0 & b_1 & 0 & 0 &\dots & 0\\[1ex]
		~-b_1~ & 0 & b_2 & 0 &\dots & 0\\[1ex]
		0 & -b_2 & 0 & b_3 &\dots & 0\\[1ex]
		0 & 0 & -b_3 &\ddots &\ddots & \vdots\\[1ex]
		\vdots & & & \ddots & \ddots & ~b_{N}~\\[1ex]
		0 & 0 & 0 & \dots &-b_{N} & 0\\
	\end{pmatrix}\,.
	\end{equation}
	A direct substitution in the corresponding Lax equation 
	$$    \frac{dL}{dt}=[R_\pm \nabla H_1(L),L]$$
	gives the open finite Toda lattice equations in Flaschka's coordinates $a_n$, $b_n$
	\begin{equation}  \label{eq:Flaschka_coord}
	\begin{cases}
		\dot{a}_1=2b_1^2\,,\qquad \dot{a}_{N+1}=-2b_{N}^2\,,\\[1.5ex]		\dot{a}_j=2(b_{j}^2-b_{j-1}^2)\,,\qquad j=2,\dots,N\,, \\[1.5ex]
            \dot{b}_j=b_j(a_{j+1}-a_j)\,,\qquad j=1,\dots,N\,.
	\end{cases}
	\end{equation} 
	
\subsection{\boldmath Open Toda chain with a skew-symmetric $r$-matrix~\cite{CauDelSin}}\label{Toda_pq}
We now present the same model for the same algebra $\mathfrak{g}=\mathfrak{sl}(N+1)$ but endowed with a different Lie dialgebra structure. This is based on the Cartan decomposition of $\mathfrak{g}$ and leads to a skew-symmetric $r$-matrix. One attractive feature of this setup, that we only illustrate for $\mathfrak{sl}(N+1)$, is that it allows for a generalisation to any finite semi-simple Lie algebra (\cite[Chapter 4]{BBT}). 
Consider the decomposition 
	\begin{equation}  
 \label{decomp_g}
		\mathfrak{g} =  \mathfrak{n}_+ \oplus \mathfrak{h} \oplus \mathfrak{n}_- \,, 
	\end{equation}
	where $\mathfrak{h}$ is the Cartan subalgebra of diagonal (traceless) matrices and $\mathfrak{n}_\pm$ the nilpotent subalgebra of strictly upper/lower triangular matrices. 
 Let $P_\pm$, $P_0$ be the projectors onto $\mathfrak{n}_\pm$ and $\mathfrak{h}$ respectively, relative to the decomposition \eqref{decomp_g} and set $R=P_+-P_-$. It can be verified that $R$ satisfies the mCYBE. Here $R_{\pm}=\pm(P_\pm+P_0/2)$ and 	
 \begin{equation}
		\mathfrak{g}_{\pm} = \text{Im}(R_{\pm}) = \mathfrak{b}_{\pm} = \mathfrak{h} \oplus \mathfrak{n}_{\pm}\,. 
	\end{equation}
We have the following action of $R_\pm$ on the elements $y \in \mathfrak{h}$ and $w_{\pm} \in \mathfrak{n}_{\pm}$,
	\begin{equation} \label{eq:R_action_BBT}
		R_{\pm} (y) = \pm \frac{1}{2}\, y \,,\qquad 
		R_{\pm} (w_{\pm}) = \pm \, w_{\pm} \,,\qquad 
		R_{\pm} (w_{\mp}) = 0 \,.
	\end{equation}
 Taking the same bilinear form as in \eqref{bilinear_form} $\langle X| Y\rangle=\text{tr}(XY)$, we see that 
 \begin{eqnarray}
 \label{adjoints}
     P_\pm^*=P_\mp\,,~~P_0^*=P_0~~\text{ so that}~~ R^*=-R\,.
 \end{eqnarray}
 Thus, we have a skew-symmetric $r$-matrix here. 
	For the related Lie groups, we have the following factorisations close to the identity, 
	\begin{equation} \label{eq:phi_BBT}
		g =  g_{+}\,g_-^{-1} \,, \qquad g_{\pm} =  W_{\pm}\,Y^{\pm 1} \,, \qquad Y\in \text{exp}(\mathfrak{h}) \,,\,\, W_{\pm}\in\text{exp}(\mathfrak{n}_{\pm}) \,. 
	\end{equation}
For $\Lambda \in \mathfrak{g}^*\simeq \mathfrak{g}$, the expression of $L$ as a coadjoint orbit of $\Lambda$ is given by
	\begin{equation}
		\begin{split} \label{eq:L_mf_toda_complete}
			L &= \text{Ad}^{R*}_{g} \cdot \Lambda = R^{*}_+( W_+\,Y\,\Lambda\,Y^{-1}\,W_+^{-1}) - R^{*}_-(W_-\,Y^{-1} \,\Lambda\,Y\,W_-^{-1}) \,.
		\end{split}
	\end{equation}
	We choose $\Lambda$ as in \eqref{eq:Lambda}, emphasising that in this case it is an element of the full $\mathfrak{g}^* \simeq \mathfrak{g}$, and $Y \in \text{exp}(\mathfrak{h})$, $W_{\pm}\in \text{exp}(\mathfrak{n}_{\pm})$ given by
	\begin{align*} 
		&\hspace{20ex}Y = \text{diag} \left( \eta_1 \,, \eta_2 \dots, \eta_{N+1} \right) \,, \qquad \det Y = 1\,, \\[2ex]
		&W_- \!=\! {\small \begin{pmatrix} 
			1 & 0  & 0 &\dots & 0\\[1.5ex]
			\omega^-_{2,1} & 1  & 0 &\dots & 0\\[1.5ex]
			\omega^-_{3,1} & \omega^-_{3,2} & 1 &\dots & 0\\[1.5ex]
			\vdots & \ddots & \ddots & \ddots & ~~0~~\\[1.5ex]
			~\omega^-_{N,1}~ & \omega^-_{N,2}  & \dots &\omega^-_{N,N-1} & 1\\
		\end{pmatrix}} , ~~ W_+ \!=\! {\small \begin{pmatrix}
			1 & \omega^+_{1, 2} & \omega^+_{1, 3}  &\dots & \omega^+_{1, N}\\[1.5ex]
			0 & 1 & \omega^+_{2, 3}  &\dots & \omega^+_{2 ,N}\\[1.5ex]
			0 & 0 & 1 &\ddots & \vdots\\[1.5ex]
			\vdots & &  & \ddots & ~\omega^+_{N-1, N}~\\[1.5ex]
			~~0~~ & 0  & \dots & 0 & 1\\
		\end{pmatrix}} .
	\end{align*}
	From \eqref{adjoints}, we deduce that $R^*_{\pm}=\pm(P_\mp+P_0/2)$ so that
	\begin{equation}
			R_{\pm}^* (y) = \pm \frac{1}{2}\, y \,,\qquad 
			R_{\pm}^* (w_{\pm}) = 0 \,,\qquad  
			R_{\pm}^* (w_{\mp}) =  \pm w_{\mp} \,,
	\end{equation}
 for $y \in \mathfrak{h}\,,~w_{\pm} \in \mathfrak{n}_{\pm}$.  Let us introduce the variables $(w_i,z_i)$, defined as  
	\begin{equation} \label{eq:variables_w_z}
		w_i = \frac{\omega^+_{i,i+1} - \omega^-_{i+1,i}}{2}\,, \qquad z_i=2\,\frac{\eta_{i+1}}{\eta_i} \,,
	\end{equation}
from which we determine the Flaschka coordinates as 
	\begin{equation} \label{eq:BBT_Flash_coord}
 \begin{cases}
     	a_i = \dfrac{w_i\,z_i-w_{i-1}\, z_{i-1}}{2} \,, \qquad i = 2, \dots, N-1\,, \\[2ex]
      a_1 = \dfrac{w_1\,z_1}{2} \,, \qquad a_{N+1} = -\dfrac{w_N\,z_N}{2} \,,\\[2ex]
      b_i = \dfrac{z_i}{2} \,,\qquad  i=1, \dots, N\,.  
 \end{cases}
\end{equation}
The evaluation of \eqref{eq:L_mf_toda_complete} in those coordinates reproduces the tridiagonal form as in \eqref{form_L_Flaschka}. One can then check that the equations \eqref{eq:Flaschka_coord} in the previous case derive from the Lax equation 
\begin{equation*}
    \partial_{t_k} L = \left[\, R_{+}(\nabla H_k(L)),L \,\right] \,, \qquad k = 1,2 \,,
\end{equation*}
where the Hamiltonians are taken as 
\begin{equation}
    H_1(L) = \text{tr}\,(L^2)\,, 
\end{equation}
and we recall that $R_+=P_+ +P_0/2$ here.
\subsection{Toda lattice from Blaszak-Marciniak~\cite{BlaMar}}
The approach of the Lie dialgebras can be extended to loop algebras as well. We will not deepen the subject, but we will just show the results for a very simple case of lattice equation, that gives the periodic Toda lattice. 

We consider the space of discrete fields $u_{i,n} \colon \mathbb{Z} \to \mathbb{R} $, $i \in \mathbb{N}$. We can introduce the algebra of the shift operators $S$ 
\begin{equation}
    L = u_{i,n}\,S^i = u_{0,n} \Lambda^0 + u_{1,n}\Lambda^1 + u_{2,n}\Lambda^2 + \dots  \,, \qquad L \in \mathfrak{g}\,,
\end{equation}
with $\Lambda = \{\delta_{i,i+1}\}_{i \ge 0}$
for shift operators defined by the action  
\begin{equation}
    S\,u_{i,n} = u_{i,n+1}
\end{equation}
and a commutation rule 
\begin{equation}
    S^k u_{i,n} = (S^k u_{i,n}) S^k \,, 
\end{equation}
which is an associative algebra. We introduce the trace on $\mathfrak{g}$  
\begin{equation}
    \text{tr}(L) = \text{tr}(u_{i,n} S^i) = \sum_{n=-\infty}^{\infty} u_{0,n}\,. 
\end{equation}
The algebra $\mathfrak{g}$ is uniquely determined by the set of coefficient fields $(\dots, u_{-1}, u_0, u_1, \dots )^{\top}$. 

We consider the decomposition of the Lie algebra $g$ as a vector space 
\begin{align}
    \mathfrak{g} = \mathfrak{g}_+ \oplus \mathfrak{g}_- \simeq \mathfrak{g}_{\ge k} \oplus \mathfrak{g}_{<k}\,, \hspace{5ex} \\
    \mathfrak{g}_{\ge k} = \sum_{i\ge k} u_i\, S^i \,, \qquad \mathfrak{g}_{< k} = \sum_{i < k} u_i\, S^i \,.
\end{align}
One question we can consider is for what values of $k$ the subspaces $\mathfrak{g}_{< k}$ and $\mathfrak{g}_{\ge k}$ are indeed subalgebras. Let us consider $L_1, L_2 \in \mathfrak{g}_{\ge k}$, i.e. 
\begin{equation}
    \begin{split}
        L_1 &= u_k S^{k} + u_{k+1} S^{k+1}+\dots\,, \quad 
        L_2 = v_k S^{k} + v_{k+1} S^{k+1}+\dots\,, \\
    \end{split}
\end{equation}
then the commutator is 
\begin{equation}
    [L_1, L_2] = [u_k,v_k] S^{2k} + \dots\,, 
\end{equation}
which is still an element of $\mathfrak{g}_{\ge k}$ if and only if 
\begin{equation}
    2k \ge k ~~ \implies ~~ k \ge 0\,. 
\end{equation}
We consider two elements $L_3, L_4 \in \mathfrak{g}_{<k}$ 
\begin{equation}
    \begin{split}
        L_3 &= \dots + u_{k-1} S^{k-1}\,, \\
        L_3 &= \dots + v_{k-1} S^{k-1}\,,  \\
    \end{split}
\end{equation}
and the commutator 
\begin{equation}
    [L_3, L_4] = \dots + [u_{k-1},v_{k-1}]S^{2k-2} 
\end{equation}
is still in $\mathfrak{g}_{<k}$ if and only if 
\begin{equation}
    2k-2 < k ~~ \implies ~~ k < 2\,.
\end{equation}
The only two cases for which $\mathfrak{g}_{\ge k}$ and $\mathfrak{g}_{<k}$ are indeed Lie subalgebras of $\mathfrak{g}$ are for $k\in\{0,1\}$. The $r$-matrix $R$ is 
\begin{equation}
    R_k = P_{\ge k} - P_{<k}\,, \qquad k\in \{0,1\}\,. 
\end{equation}
It is possible to consider $m$-component reduction of $L$ expressed as 
\begin{equation}
    L = u_{\ell+m} S^{\ell+m} + u_{\ell+m-1}S^{\ell+m-1}+ \dots + u_{\ell}S^{\ell}\,,\qquad -m < \ell \le -1\,,
\end{equation}
and for $k=0$ we can set $u_{\ell+m}=1$, while for  $k=1$ we can set $u_{\ell}=1$. 

In particular, for $k=0$, $\ell=1$ we have (identifying the Flaschka coordinates)
\begin{equation}
    L = u_{-1}S^{-1} + u_0+S = a_nS^{-1}+b_n+S\,, 
\end{equation}
which enter the Lax equation 
\begin{equation}
    \frac{dL}{dt} = [(L)_{\ge0},L]\,.
\end{equation}
The left hand side of this is
\begin{equation*}
    \frac{dL}{dt} = \dot{a}_n\,S^{-1} + \dot{b}_n \,,   
\end{equation*}
and the right hand side 
\begin{equation}
\begin{split}
     [(L)_{\ge0},L] &= [b_n+S, a_nS^{-1}+b_n+S ] \\
     &= b_n a_n S^{-1}-a_n b_{n-1}S^{-1} + a_{n+1}SS^{-1} - a_n S^{-1}S \,. 
\end{split}
\end{equation}
Comparing the two sides we consider the powers of the shift operators $S^{-1}$, $S^0=I$ and we obtain the (periodic) Toda equations for a different parametrisation. 


\bibliographystyle{amsplain}

\end{document}